\documentclass[twocol]{ametsocV6.1}

\title{An Earth-System-Oriented View of the S2S Predictability of North American Weather Regimes}

\authors{Jhayron S. Pérez-Carrasquilla\aff{a}\correspondingauthor{Jhayron S. Pérez-Carrasquilla, jhayron@umd.edu}
    and Maria J. Molina\aff{a} 
}

\affiliation{
    \aff{a}{Department of Atmospheric and Oceanic Science, University of Maryland, College Park, Maryland, USA}\\
}

\abstract{It is largely agreed that subseasonal-to-seasonal (S2S) predictability arises from the atmospheric initial state during early lead times and from the land and ocean during intermediate and late lead times. We test this hypothesis for the large-scale mid-latitude atmosphere by training numerous XGBoost models to predict weather regimes (WRs) over North America at 1-to-8-week lead times. Each model uses a different predictor from one Earth system component (atmosphere, ocean, or land) sourced from reanalysis. According to the models, the atmosphere provides more predictability during the first two forecast weeks, and the three components performed similarly afterward. However, the skill and sources of predictability are highly dependent on the season and target WR. Our results show greater WR predictability in fall and winter, particularly for the Pacific Trough and Pacific Ridge regimes, driven primarily by the ocean (e.g., El Niño-Southern Oscillation and sea ice). For the Pacific Ridge in winter, the stratosphere also contributes significantly to predictability across most S2S lead times. Additionally, the initial large-scale tropospheric structure (encompassing the tropics and extra-tropics, e.g., Madden-Julian Oscillation) and soil conditions play a relevant role—most notably for the Greenland High regime in winter. This study highlights previously identified sources of predictability for the large-scale atmosphere and gives insight into new sources for future study. Given how closely linked WRs are to surface precipitation and temperature anomalies, storm tracks, and extreme events, the study results contribute to improving S2S prediction of surface weather.}

\begin{document}

\maketitle

\statement
Recurrent and persistent large-scale atmospheric patterns are referred to as weather regimes. Algorithms can be trained to predict weather regimes over North America up to eight weeks into the future. Contributions to prediction skill can originate from different parts of the Earth system, such as stratospheric wind anomalies, sub-surface ocean heat content, and soil water content. In this study, we found that contributions to the predictability of weather regimes can vary in importance depending on the lead time, season, and predicted weather regime. These results show that predictability contributions can vary more than previously documented and deepen our understanding of sources of predictability, which can help improve forecasts extending beyond two weeks for the benefit of society and decision-makers.

\section{Introduction}

The subseasonal-to-seasonal timescale (S2S; two weeks to two months) is often called a ``predictability desert'' \citep{vitart2012subseasonal}. Despite research advances \citep{becker2022decade,sengupta2022advances}, S2S forecasting remains challenging due to the chaotic nature of the atmosphere, imprecise knowledge of initial conditions, and multiple interactions and feedbacks among Earth system components \citep{lorenz1963deterministic,white2017potential,pegion2019subseasonal,merryfield2020current}. For example, atmospheric features evolve more quickly than the land and ocean, but they partly determine future land and ocean conditions. The resulting oceanic and land states also affect the subsequent atmospheric state via vertical fluxes of heat and moisture \citep{CircumglobalResponsetoPrescribedSoilMoistureoverNorthAmerica}. Recently, machine learning (ML) has shown potential for S2S forecasting given its ability to capture nonlinear and complex relationships \citep[e.g.,][]{molina2023review,chen2024machine}. Here, we leverage ML to explore Earth system component contributions to S2S predictability \citep[e.g.,][]{mayer2021subseasonal,molina2023subseasonal}.

Aiming to make the forecasting problem more general while maintaining it relevant, we focus on the predictability of the large-scale atmosphere instead of focusing of specific surface weather anomalies. K-means clustering (i.e., an unsupervised ML method) has been used to identify large-scale weather or circulation regimes, which are planetary-scale flow configurations that are recurrent, persistent, and quasi-stationary \citep{michelangeli1995weather, reinhold1982dynamics,straus2007,molina2023subseasonal,lee2023new,lee2024}. These regimes persist longer and have a slower evolution than small-scale atmospheric events, some of which may be stochastic (e.g., thunderstorms), which makes them useful for S2S atmospheric prediction \citep{jennrich2024}. Regimes have an imprint on surface precipitation and temperature anomalies by creating persistent dynamic or thermodynamic conditions that lead to subsidence or convection, and by guiding the track of extratropical cyclones \citep[e.g.,][]{vigaud2018predictability}. They are also related to the occurrence of extreme events \citep{lee2023new,jennrich2024}. \cite{molina2023subseasonal} found that upstream outgoing longwave radiation (OLR) and sea surface temperatures (SSTs) help skillfully predict regimes several weeks in advance during the cold season. Also, for the boreal winter, \cite{jennrich2024} demonstrated that precipitation and temperature forecasts could improve by using WR forecasts and their respective composited anomalies instead of the raw output fields from the models. Finally, \cite{lee2023new} redefined North American regimes, allowing them to be used for the whole year instead of just for the cold season.

Beyond helping identify persistent and recurrent patterns, ML can be used to forecast the weather regimes (WRs). Many recent developments, mainly using neural networks, have demonstrated that the performance of ML is comparable to state-of-the-art numerical models at S2S lead times \citep{weyn2021sub,weyn2021can,bi2023accurate,lang2024aifsecmwfsdatadriven,chen2024machine}. However, neural networks can have a large number of hyperparameters, be prone to overfitting, and take substantial computing to train \citep[e.g.,][]{schmitt2022deep}. On the other hand, tree-based algorithms can still decipher nonlinear patterns among training data \citep{herman2018money}, and have the capacity to achieve similar performance as neural networks at S2S lead times at comparably lower computational costs during training \citep[e.g.,][]{vitart2022outcomes}. Tree-based models include Extreme Gradient Boosting \citep[XGBoost;][]{friedman2001greedy}, which has been used frequently during recent years to forecast a wide range of phenomena \citep{molina2023review}, such as lightning \citep{mostajabi2019nowcasting}, daily precipitation \citep{dong2023enhancing}, and seasonal temperature \citep{qian2020machine}. XGBoost presents several advantages over other tree-based methods because it can better handle imbalanced datasets, has less likelihood of overfitting, and exhibits better detection of non-linear patterns \citep{fatima2023xgboost}. These advantages are mainly due to a more flexible and adaptable architecture and the dynamic adjustment of hyperparameters during training.

Given the intertwined temporal evolution of the atmosphere, land, and ocean, all three components are relevant for predictability at the S2S timescale \citep{alexander1992midlatitude,koster2010contribution,TheSecondPhaseoftheGlobalLandAtmosphereCouplingExperimentSoilMoistureContributionstoSubseasonalForecastSkill,guo2012rebound,hartmann2015pacific}. It has been largely understood that predictability arises from the atmospheric initial state at lead times of less than two weeks, in contrast to the land and ocean which provide predictability during longer lead times \citep{meehl2021initialized}. However, \cite{richter2024quantifying} demonstrated that contributions to predictability from Earth system components may be more complicated, with predictability from the atmosphere playing a greater role during S2S lead times. The initial conditions of the atmosphere, land, and ocean sometimes provide ``forecasts of opportunity'' \citep{mariotti2020windows}, some of which can manifest during certain states of the Madden-Julian Oscillation \citep[MJO;][]{robertson2018sub}, sudden stratospheric warmings \citep{kidston2015stratospheric}, and the El Niño-Southern Oscillation \citep[ENSO;][]{PredictabilityofWeek34AverageTemperatureandPrecipitationovertheContiguousUnitedStates}. Considering the theoretical and technical advances mentioned above, this study aims to quantify relative initial state contributions to weather regime predictability from various Earth system components at S2S lead times using the XGBoost algorithm \citep{chen2016xgboost}. 

\section{Data and Methods}\label{sec:methods}

\subsection{Weather Regimes Computation and Output of the Models}\label{sec:methods_wr}

\begin{figure*}[h]
 \centerline{\includegraphics[width=39pc]{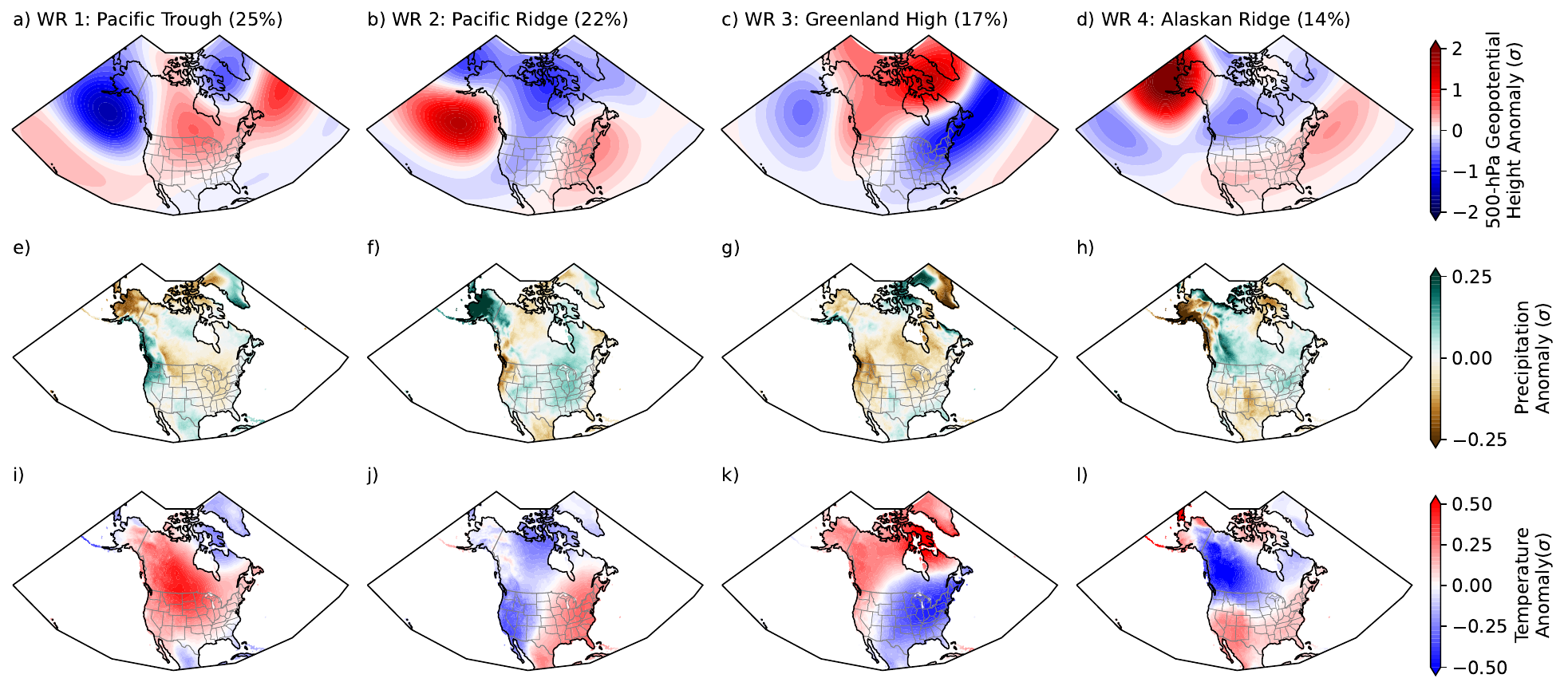}}
  \caption{The upper column (a-d) shows composites of detrended standardized ERA5 500-hPa geopotential height anomalies from 1981 to 2020 for the four weather regimes (WRs) identified. The daily frequency of each weather regime is shown in the subplot titles. The `` No WR" class is not shown. The middle and bottom columns show the composited standardized daily precipitation (e-h) and temperature (i-l) anomalies for each one of the four weather regimes sourced from ERA5.}
  \label{f1}
\end{figure*}

Hourly geopotential height at 500-hPa (Z500) on an approximately 31-km grid was sourced from the ECMWF fifth generation reanalysis \citep[ERA5;][]{hersbach2020era5} to compute WR classes. Our study was constrained to 1981-2020 due to the limited availability of ocean data below the sea surface from a different product. Following the \cite{lee2023new} framework, we computed year-round daily WRs. To do this, we first computed the daily average for each grid cell over the region shown in Figure \ref{f1} (20$^\circ$N to 80$^\circ$N and 180$^\circ$W to 30$^\circ$W). Then, we calculated daily anomalies by subtracting the respective grid cell's multi-year daily average from 1981 to 2020. We applied a 10-day low-pass Fourier filter (across the time dimension) on the anomalies to emphasize synoptic-scale variability. Subsequently, we detrended the data by subtracting the daily area-averaged (cosine-latitude weighted) linear trend of the Z500 field over the domain (5.73 m/decade), and we divided the data by the multi-year daily standard deviation averaged over the region to obtain standardized anomalies. Both the annual cycle of the mean and the standard deviation were smoothed using a 60-day window. Although considering the full period for computing the climatology is not the standard procedure when training ML models, we emphasize that our objective is to gain insight into the sources of predictability and not to develop a realistic prediction system. Thus, this represents a minimal study limitation.

After obtaining the detrended standardized anomalies at a daily temporal resolution, the methodology described by \cite{molina2023subseasonal} and \cite{lee2023new} was replicated. First, we extracted the 12 leading principal components (PCs) from the Z500 fields. Then, we identified the WRs by applying k-means clustering with four centroids over the time series of these 12 PCs \citep{michelangeli1995weather}. Following \cite{lee2023new}, we created a fifth class called ``No WR," which corresponds to samples that are closer to the origin than to any of the four WR centroids; that is, the PC scores are closer to the (0, 0, 0, ... 0) point on the PC space. Additionally, aiming to achieve higher quality in the clustering, we included in the ``No-WR class" the 10\% of samples with the highest distance to their corresponding centroids in the PC space. 

The average Z500 detrended standardized anomalies associated with the four identified WR classes are shown in Figure \ref{f1}. Consistent with nomenclature from past studies \citep[e.g.,][]{lee2019wintertime,robertson2020toward,molina2023subseasonal,lee2023new}, the identified large-scale WRs were named Pacific Trough (PT), Pacific Ridge (PR), Greenland High (GH), and Alaskan Ridge (AR). As shown in previous studies \citep{molina2023subseasonal,lee2023new,jennrich2024} and in Figure \ref{f1}e-l, each weather regime is associated with specific surface temperature and precipitation anomalies. The Pacific Trough regime is associated with increased precipitation over the U.S. Pacific Northwest, dry conditions over the Midwest, and increased temperatures over Central U.S. and Canada. The Pacific Ridge regime is linked to increased precipitation over Alaska and the Midwest, dry conditions over the U.S. Pacific Northwest, colder-than-average temperatures over the U.S. West Coast, and vice-versa for the U.S. East Coast. The Greenland High regime is associated with dry conditions over the U.S. Pacific Northwest and Eastern Canada, with cold temperature anomalies over the U.S. East Coast and warm temperature anomalies over Western Canada. Lastly, the Alaskan Ridge regime presents wet anomalies over British Columbia, dry conditions over Alaska and the U.S. South East, cold anomalies over Western Canada, and warm anomalies over the Western U.S. Therefore, accurate prediction of the WRs has direct implications for these specific regions. The objective of the models described below is to predict the WR present during the majority of the days of the target week. The weeks with no predominant weather regime were labeled as ``No WR" weeks.

\subsection{Sources of Predictability and Input Data for the Models}\label{sec:methods_data}

Disregarding external forcings, the primary sources of S2S predictability can be classified into two categories: i) the state of recurring or quasi-oscillatory patterns, and ii) anomalies in the initial state of one of the components of the Earth system whose typical timescale of evolution (i.e., persistence time) is similar to the target forecast \citep{board2016next}. In this study, we start from the (imperfect) assumption that most of the information corresponding to those two categories is contained within a limited set of variables from three Earth system components (ocean, land, and atmosphere). Additionally, we assume that the XGBoost models, in conjunction with a sufficiently large sample size, are skillful enough to at least partially represent the physical laws that determine the spatiotemporal behavior of the processes affecting the future state of the atmosphere within the S2S temporal horizon.

Each model within this study is trained separately using one variable from an Earth system component. For the atmosphere and land, we used data from ERA5 and ERA5-Land \citep{munoz2021era5} reanalyses, respectively. For the ocean, we used data from the Simple Ocean Data Assimilation reanalysis \citep[SODA;][]{carton2018soda3}. Taking into account various known sources of predictability originating from tropical and extratropical regions \citep[e.g.,][]{stan2017review, tseng2018prediction, lin2019tropical}, the spatial domain of all input variables span all longitudes (360$^\circ$ around the prime meridian) and latitudes from 30$^\circ$S to 90$^\circ$N. For the input variables, all fields were re-gridded onto a 2$^\circ$ horizontal grid using nearest neighbor interpolation. Finer spatial resolutions were tested without a notable increase in the skill of the models. In the following subsections, we describe the potential sources of predictability we aim to represent and the corresponding input variables.

All input variables were transformed into weekly detrended standardized anomalies following the steps described in Subsection \ref{sec:methods_wr} and averaged each week. However, instead of removing the area average trend, each pixel was detrended separately, and no Fourier filtering was performed. After obtaining the weekly detrended standardized anomalies by subtracting the daily climatology and dividing by the standard deviation of each grid cell, we smoothed the fields spatially using a 3x3 moving kernel. Apart from the models trained separately to assess predictability from each individual variable, three other models were trained using atmosphere-only, land-only, and ocean-only variables as inputs. In addition to the fields of standardized anomalies, we added the sine and cosine of the day of the year as another input to the models to improve the representation of seasonal differences in Earth system processes.

\subsubsection{Atmospheric variables}\label{sec:methods_data_atm}

Four atmospheric variables were used in our study, with three from the troposphere and one from the stratosphere (Table \ref{t1}). The variables from the troposphere are Z500, outgoing long-wave radiation (OLR), and 200-hPa zonal wind (U200). Z500 describes the mid-latitude atmosphere, including the large-scale atmospheric circulation \citep{cheng1993cluster}, North-Atlantic oscillation \citep[NAO;][]{wallace1981teleconnections,davini2012coupling}, planetary- and synoptic-scale waves \citep{blackmon1976climatological}, and blocking episodes \citep{tibaldi1990operational, scherrer2006two}. OLR provides information about the strength and spatial coverage of convective clusters, which are related to tropical-extratropical teleconnections \citep{stan2017review} from the MJO \citep{hendon1994life, wheeler1999convectively, wheeler2004all}, ENSO \citep{yulaeva1994signature}, and Indian Ocean dipole \citep[IOD;][]{saji1999dipole}. U200 provides information about the shape and location of the polar and subtropical jet streams \citep{krishnamurti1961subtropical,palmen1948distribution,strong2007winter}, in addition to the coupling of the Walker circulation to the phase of ocean-related phenomena, like ENSO or IOD \citep{philander1983nino,saji1999dipole}.

\begin{table*}
\caption{The four atmospheric variables of the Earth system used for training the ML models sourced from ERA5. The relevance for S2S forecasting is also indicated.}\label{t1}
\begin{center}
\begin{tabular}{cc}
\topline
\textbf{Variable} & \textbf{Relevance to S2S forecasting}\\
\midline
500hPa geopotential height (Z500) & Mid-latitude atmosphere (e.g., weather regimes, North-Atlantic oscillation, \\
& planetary Rossby waves, and blocking episodes).\\
Outgoing longwave radiation (OLR) & Clusters of convection and large-scale cloud patterns.\\
10hPa zonal wind (U10) & Quasi-biennal oscillation and sudden stratospheric warming events.\\
200hPa zonal wind (U200) & Strength and structure of the walker cell, jet streams, and Indian Ocean dipole.\\
\botline
\end{tabular}
\end{center}
\end{table*}

The stratosphere is a valuable source of S2S predictability given its slowly varying circulation anomalies \citep{board2016next}. During boreal winter, the stratosphere can interact with the upper troposphere and modify the tropospheric circulation, along with the onset and evolution of WRs \citep{baldwin2001stratospheric,baldwin2003stratospheric,gerber2009stratospheric,domeisen2020role}. Two primary stratospheric mechanisms contributing to mid-latitude predictability have been previously identified, the quasi-biennial oscillation \citep[QBO;][]{baldwin2001quasi} and the stratospheric polar vortex \citep{gerber2009stratospheric}. The QBO is an easterly-to-westerly reversal of tropical stratospheric winds partly driven by waves originating from the troposphere that can affect the strength of the stratospheric polar vortex \citep{baldwin2001quasi}. Changes in the stratospheric polar vortex can result in large-scale tropospheric circulation anomalies with a lag of about one month \citep{thompson2000annular,baldwin2001stratospheric}. Rapid slowdowns of the stratospheric polar vortex are frequently accompanied by sudden stratospheric warmings that can affect the Northern Annular Mode and contribute to an equatorward shift of the tropospheric midlatitude jet \citep{gerber2009stratospheric,domeisen2020role}. The zonal wind at 10-hPa (U10) is the fourth atmospheric variable we consider because it can help diagnose the QBO phase and the strength of the stratospheric polar vortex \citep{baldwin2001quasi,gerber2009stratospheric}.

\subsubsection{Oceanic variables}\label{sec:methods_data_ocean}

The ocean is relevant for S2S prediction because it provides information about natural modes of climate variability and it can influence the overlying atmosphere \citep{board2016next}. Perturbations of the ocean's mean state, particularly within the tropics, can induce organized large-scale convection, modifying the heating distribution through the atmosphere and producing temperature and precipitation anomalies. Organized convection can also change the extratropical circulation due to propagation via Rossby wave-trains \citep{bjerknes1969atmospheric,horel1981planetary,trenberth1998progress}. ENSO is one of the primary modulators of long-lived tropical signals, with positive or negative SST anomalies over the eastern-to-central equatorial Pacific characterizing its warm and cold phases respectively \citep{philander1983nino}. 

The atmosphere can also respond to extratropical ocean features \citep{zhou2019atmospheric}, such as eddy-mediated processes, which can generate near-surface atmospheric baroclinicity and modify the spatial characteristics of overlying storm tracks \citep{kwon2010role}. Another example is a linear thermodynamic interaction, where geopotential heights are modulated above the ocean anomaly \citep{kushnir2002atmospheric}. Although the atmospheric variance associated with extratropical SST anomalies is modest compared to internal atmospheric variability, extratropical SST anomalies can persist for months to years \citep{chelton2010coupled} and the atmosphere can respond within a few months \citep{ferreira2005transient,deser2007transient}. 

Ocean data was sourced from SODA reanalysis (Table \ref{t1b}). To capture information about subsurface conditions that may reach the surface via mixing \citep{alexander1992midlatitude}, we used ocean heat content (OHC) integrated from the surface to different depths (50m, 100m, 200m, 300m, and 700m). Sea surface height (SSH) and mixed layer depth (MLD) were also used, which provide information about the upper-ocean circulation, anomalous upwelling or downwelling, and the size, depth, and location of eddies \citep{chelton2011global,kurczyn2012mesoscale,uchoa2023brazil}. 

\begin{table*}
\caption{The 10 oceanic variables from the Earth system used for training the ML models sourced from SODA. The relevance for S2S forecasting is also indicated.}\label{t1b}
\begin{center}
\begin{tabular}{cc}
\topline
\textbf{Variable} & \textbf{Relevance to S2S forecasting} \\
\midline
Sea surface temperature (SST) & ENSO, storage of anomalous surface heat, and eddies.\\
Ocean heat content (OHC) integrated from the surface & Magnitude and persistence of anomalous heat.\\
to a 50m, 100m, 200m, 300m, or 700m depth.&\\
Mixed-layer depth (MLD)&Anomalous upwelling or downwelling, and upper-ocean circulation structure.\\
Sea-surface height (SSH)&Anomalous upwelling or downwelling, and upper-ocean circulation structure.\\
Sea ice thickness (IT) &Anomalous changes of heat and moisture fluxes from the ocean to the atmosphere,\\
& and changes in albedo.\\
Sea ice concentration (IC) & Anomalous changes of heat and moisture fluxes from the ocean to the atmosphere,\\
& and changes in albedo.\\
\botline
\end{tabular}
\end{center}
\end{table*}

Sea ice thickness and concentration anomalies were also used, which are generally slow-evolving and describe sea ice cover (Table \ref{t1b}). Sea ice can significantly reduce heat and moisture fluxes from the ocean to the atmosphere, modifying the surface albedo and increasing how much incoming radiation is reflected \citep{board2016next}. The presence and persistence of sea ice can also affect the paths of atmospheric lows \citep{balmaseda2010impact,screen2011dramatic}.

\subsubsection{Land variables}\label{sec:methods_data_land}

Soil water content can influence surface energy budgets, contributing to anomalies in temperature and precipitation, and is therefore another source of S2S predictability \citep{koster2004regions,koster2010contribution,guo2011land,koster2015interactive,roundy2015attribution,thomas2016influence}. ERA5-Land, based on the Tiled ECMWF Scheme for Surface Exchanges over Land \citep[TESSEL;][]{balsamo2009revised} and forced with ERA5 atmospheric fields, provides temperature and water content for four layers down to a depth of 2.89m. Soil water content (SWC) and soil temperature (ST) integrated from the surface to the depth of each one of the four reanalysis layers (7cm, 28cm, 1m, and 2.89m) are used as input variables (Table \ref{t1c}). Terrestrial snow is another source of S2S predictability given its radiative effects and liquid water that is released during melting \citep{peings2011snow,jeong2013impacts,orsolini2013impact,thomas2016influence}. Thus, we used snow depth (SD) as an input variable for some land-only ML models.

\begin{table*}[h]
\caption{The nine land variables of the Earth system used for training ML models sourced from ERA5-Land. The relevance for S2S forecasting is also indicated.}\label{t1c}
\begin{center}
\begin{tabular}{cc}
\topline
\textbf{Variable} & \textbf{Relevance to S2S forecasting} \\
\midline
Soil water content (SWC) integrated from the surface & Storage of surface water, and proxy for soil and\\
to a 7cm, 28cm, 1m, or 2.89m depth. & vegetation characteristics. \\
Soil temperature (STL) averaged from the surface & Proxy for potential evaporation and proxy for soil\\
to a 7cm, 28cm, 1m, or 2.89m depth.& and vegetation characteristics. \\
Snow depth (SD) & Storage of surface water and influence on surface energy budgets.\\
\botline
\end{tabular}
\end{center}
\end{table*}

% \newpage
\subsection{Forecasting framework}\label{sec:methods_framework}

Numerous XGBoost models were independently trained to predict weekly weather regime (WR) classes over North America. We trained each model with a different input variable to assess the contribution of the respective predictor at different lead times, ranging from 1 to 8 weeks into the future. Weekly averaged data was used with weeks delineated on Mondays and Thursdays; some overlap in the days of the week was permitted for the weekly averages in exchange for a doubling of sample size. For any weekly forecast that started on Monday ($t\,=\,0$), the input ``week 0'' contained days from the previous Tuesday and finished on that respective Monday ($t\,=\,-6,\,...\, 0$), while the first lead time of the forecast (i.e., week 1) included days from the next Tuesday to the following Monday ($t\,=\,1,\,...\, 7$), and so on for weeks 2 through 8. Likewise, for weekly forecasts starting on a Thursday ($t\,=\,0$), the weekly input contained days from the previous Friday ending that Thursday.

\subsection{XGBoost hyperparameters}\label{sec:methods_architecture}

We performed a Bayesian optimization \citep{snoek2012practical} of hyperparameters for each ML model to obtain the combinations that produced the best possible performance within the limits of our study. We emphasize that our objective is not to obtain the best possible S2S forecasts but to assess predictability contributions from different variables and Earth system components. Given the potential presence of certain states of modes of climate variability, a model may perform better when evaluated in one particular period than another. To prevent overfitting to such patterns, we used k-fold cross-validation to find the best combination of hyperparameters and cross-testing to obtain performance metrics representative of the whole study period. The ``cross-validation and cross-testing" approach has been shown to remove biases when assessing performance while also taking advantage of all samples within the dataset \citep{korjus2016efficient}. The hyperparameter search is performed independently for each variable and lead time.

We employed cross-validation and cross-testing in the following manner for each one of the variables. The 40-year dataset was divided into four decades (1981-1990, 1991-2000, 2001-2010, and 2011-2020). For cross-testing, we denoted a test subset as one of the four decades. A model predicting a test subset was then trained using the remaining three decades of data. Before training on the remaining three decades, we applied three-fold cross-validation, where each one of the decades served as one ``fold.'' Three-fold cross-validation allowed us to obtain the combination of hyperparameters that provided good skill across each held out fold, which involved trying multiple combinations of hyperparameters and training iteratively on two of the three decades. Finally, the combination of hyperparameters yielding the best average performance over the three decades was chosen to predict the respective test subset, and this process was repeated three more times to obtain performance statistics on the four test subsets.

Ten hyperparameters were assessed during optimization (Table \ref{t2}), including those related to the complexity of the trees (e.g., maximum tree depth) and to prevent overfitting (e.g., L1 and L2 regularization terms). Additionally, we assessed the impact of adding class weights to control for WR class imbalance. The Bayesian algorithm \citep{snoek2012practical} used to perform the hyperparameter search approximates a probabilistic function to the hyperparameter field to make iterative changes in a direction that improves the objective function. We used the F1 score as the objective function, which is similar to ``threat score'' and ``critical success index,'' and focuses on the fraction of observed and forecast events that were correctly predicted. The advantage of the F1 score over other objectives lies in that it neglects ``true negative'' cases that can artificially inflate skill when a class imbalance exists. The F1 score is computed as 

\begin{equation}
\text{F1} = \frac{2 \times \text{Precision} \times \text{Recall}}{\text{Precision} + \text{Recall}},
\end{equation}

\noindent where 

\begin{equation}
\text{Precision} = \frac{\sum_{c} \text{True Positives}_{c}}{\sum_{c} (\text{True Positives}_{c} + \text{False Positives}_{c})}, 
\end{equation}

\begin{equation}
\text{Recall} = \frac{\sum_{c} \text{True Positives}_{c}}{\sum_{c} (\text{True Positives}_{c} + \text{False Negatives}_{c})}, 
\end{equation}

\noindent and $c$ represents each of the 5 classes.

60 initial trials with random hyperparameter combinations were run to provide an initial performance baseline and then 15 subsequent iterations were performed using Bayesian optimization to find the best combination. The Bayesian algorithm was executed using the Expected Improvement method \citep{frazier2018tutorialbayesianoptimization} with $\xi = 0$ for prioritizing exploitation after the random search (larger $\xi$ values produced no improvement in performance). The number of estimators (trees or boosting rounds) in the XGBoost algorithm was 20 for each model. Trials with a higher number of trees were done without notable improvement in the models' skill. Table \ref{t2} shows the optimal hyperparameters identified during the search. The loss function used by the XGBoost algorithm is categorical cross-entropy.

\begin{table*}[h]
\caption{Set of XGBoost hyperparameters explored during the Bayesian optimization.}\label{t2}
\begin{center}
\begin{tabular}{ccccrrcrc}
\topline
\textbf{Hyperparameter} & \textbf{Search range} & \textbf{Optimal values}\\
\midline
 Maximum depth of a tree & 2-20 & 12-14\\
 Minimum sum of instance weight needed in a child & 1-20 & 10-12\\
 Percentage of samples used for each tree construction & 0.7-0.9 & 0.8-1\\
 Percentage of features used for each tree construction & 0.6-0.9 & 0-0.25\\
 Percentage of features used for each split & 0.75-1 & 0.83-0.88\\
 Learning rate & 0.0001-0.3 & 0.003-0.015\\
 Minimum loss reduction required to make a further partition on a leaf node of a tree (gamma) & 0-5 & 0-4\\
 L1 regularization term on weights & 4-40 & 6-10\\
 L2 regularization term on weights & 1-316 & 1.5-40\\
 Exponential term for the weights to reduce class imbalance & 0-1 & 0.6-0.8\\
\botline
\end{tabular}
\end{center}
\end{table*}

\section{Results}\label{sec:results}

\subsection{Sources of predictability based on ML model skill}\label{subsec:acc}

\begin{figure*}[h]
 \centerline{\includegraphics[width=33pc]{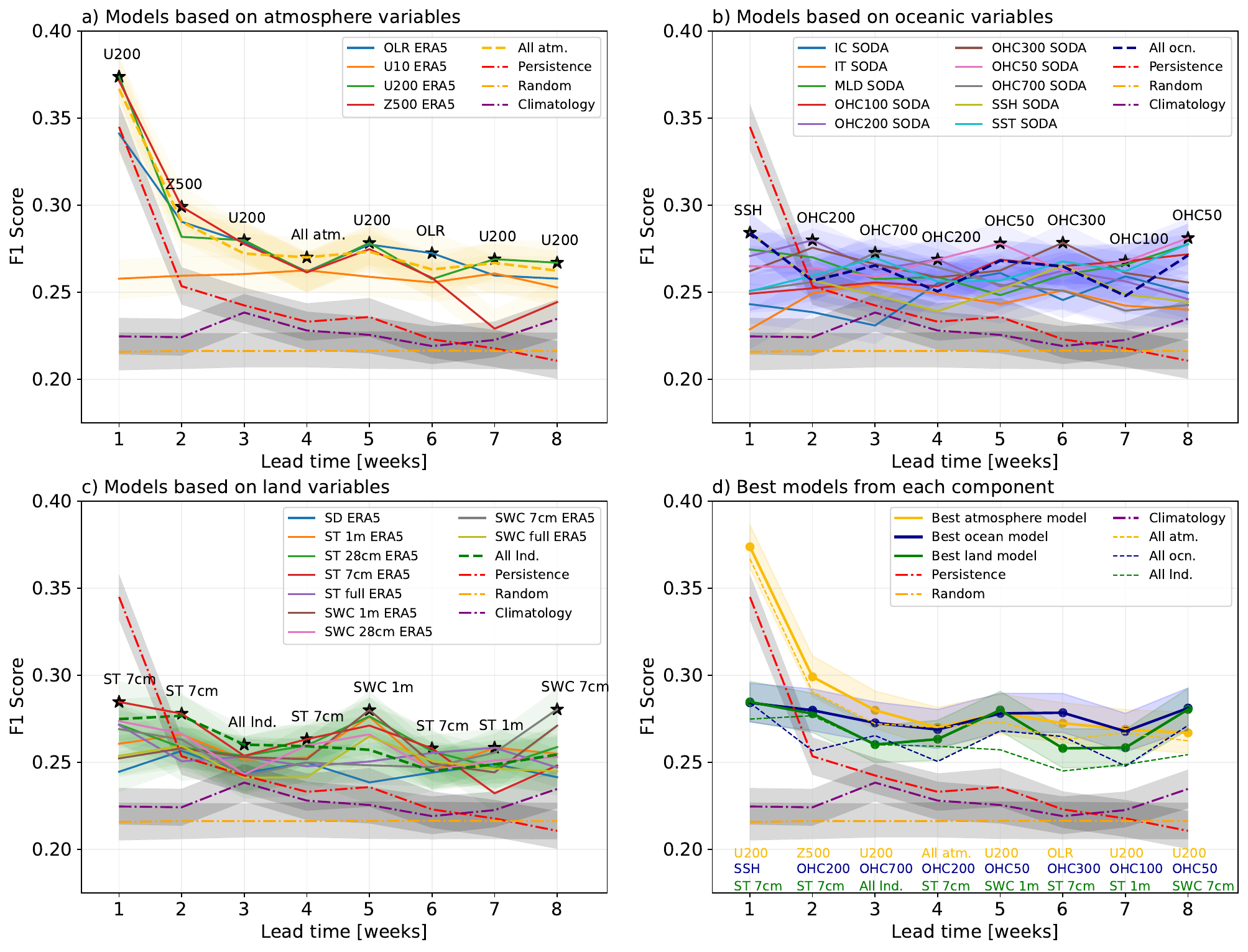}}
  \caption{Predictability from the a) atmosphere, b) land, and c) ocean indicated using F1 score as a function of lead time. The red dash-dot line shows the skill of the persistence forecast, the orange dash-dot line shows the skill of a random guess based on the annual WR climatology, and the purple dash-dot line shows the skill of a random-guess forecast based on the seasonal WR climatology. The yellow, blue, and green dashed lines show the performance of the models trained using all the predictors from the respective Earth system component, and the star markers indicate the predictor that provided, on average, more predictability at the respective lead time. d) F1 score of the best-performing model for each lead time and component, the dashed lines from a, b, and c, are also plotted. The shading in the four plots shows the 5th-95th percentile range of the F1 score from a 1,000-member bootstrap with resampling. The gray shading indicates the uncertainty associated with the benchmarks. Little or no overlap of the model and benchmark uncertainty ranges indicates meaningful predictive skill. The colored text in d) indicates the model with the best F1 score for each component at each lead time.}
  \label{f2}
\end{figure*}

Plots a, b, and c in Figure \ref{f2} show the F1 score as a function of lead time for each ML model trained using atmospheric, oceanic, or land predictors. To determine whether or not a certain variable (or set of variables) provides meaningful predictability, we compared the skill of its respective model with the skill of three benchmarks: WR persistence (red dash-dot line), random guess based on the annual WR climatology (orange dash-dot line), and random guess based on the seasonal WR climatology (purple dash-dot line). Figure \ref{f2} also shows the 5th-95th percentile ranges of the F1 score computed for each benchmark or model using a 1,000-member bootstrap with resampling. The gray shading corresponds to the benchmarks, and the gold (atmosphere), blue (ocean), and green (land) shading corresponds to different models from each component.

\begin{figure}[h]
 \centerline{\includegraphics[width=19pc]{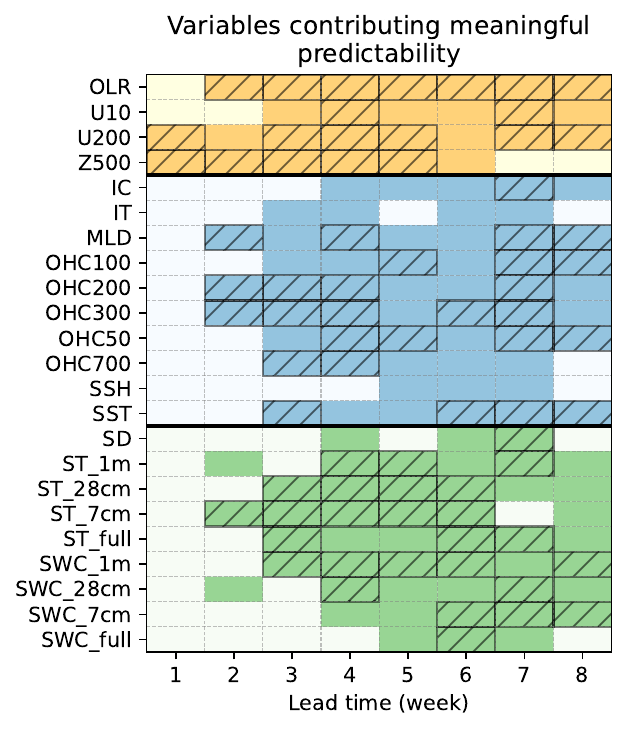}}
  \caption{Variables that contribute meaningful predictability from the atmosphere (yellow), ocean (blue), and land (green) during the different lead times. Darker shading than the background indicates that the F1 score of the model based on the respective variable was higher than the three benchmarks (and outside their confidence interval). The hatching indicates that the F1 score of the model was within the confidence interval of the best model from its respective component, indicating relatively high predictability contributions.}
  \label{f3}
\end{figure}

Meaningful predictability is obtained by a model if its F1 score is above the confidence interval of the benchmarks, with the null hypothesis being that the observed skill could be obtained by random chance with one of the three baselines. Additionally, in each one of the three plots, the dashed lines (gold for a, blue for b, and green for c) show the F1 score associated with the models trained using all the variables from each component. Figure \ref{f2}(a,b,c) also contains annotations indicating the model that presented the best skill for each lead time and each component. As observed in Figure \ref{f2}, there is high uncertainty regarding which model is the best for each component at every lead time (i.e., there is substantial overlap of model spread). Figure \ref{f3} shows the variables whose models displayed predictability and highlights the models that performed best during each lead time. It's worth noting that persistence presents skill above the other two benchmarks during weeks 1 and 2 of forecast, remaining as a significant source of atmosphere-based predictability, which highlights the usefulness of the WR framework for S2S forecasting.

For the atmosphere-only models (Figure \ref{f2}a), predictability from most of the predictors and the joint model is higher during earlier lead times (weeks 1-4) and then plateaus. The exceptions for this are the U10, which shows nearly constant skill across lead times, and Z500, which decreases continuously and becomes null after week 6. Most of the atmosphere-based models show increased predictability relative to the benchmarks in weeks 2-8 (see Figure \ref{f3}). We also note that skill was not necessarily improved when using all predictors from the atmosphere, potentially due to competing predictive signals or limitations associated with partially correlated predictors. This is also true for the other two Earth system components (Figure \ref{f2}b-c).

The predictability coming from the ocean shows nearly constant skill across lead times, as expected due to its slow evolution, with improvements over the benchmarks after week 2. Models based on OHC perform best during most lead times, and SSH, IC, and IT present predictability during fewer weeks (see Figure \ref{f3}). Although SST contributes predictability during most lead times, the performance was, on average, lower than the OHC-based models, highlighting the importance of considering subsurface processes to extract the best possible predictive skill from the ocean.

For land-based predictability the skill of the models is approximately constant across lead times. Skill (although slightly lower on average than the atmosphere and ocean) can be extracted from the land after week 3, coming from several ST- and SWC-based models (see Figure \ref{f3}). These results underscore the importance of continuing to improve the coverage of soil in situ observations \citep[e.g.,][]{hasan2020impact}. Snow depth offers limited skill compared to the other predictors. However, it's worth noting that these results consider all seasons, and some of the processes exhibit high seasonality. Therefore, a variable or component may provide skill only during certain seasons and not appear important in the year-round  analysis. Thus, we stratify results by season in the following subsection (Sect. \ref{sec:results}\ref{subsec:seasonal_acc}).

Figure \ref{f2}d compares the best-performing ML models across Earth system components. Results indicate that the atmosphere provides comparatively greater predictability at 1-2 week lead times compared to the land and ocean. Our results in Figure \ref{f2}d, although not directly comparable, are similar to those presented in Figure 1 by \cite{richter2024quantifying}, where during weeks 3-8, the predictability coming from the three components is of similar magnitude. In our case, only subtle differences are found in weeks 3 and 6, where the best atmosphere-based and ocean-based models, respectively, show higher skill than the best land-based models. Although differences among models are subtle, the three Earth system components provide valuable predictability relative to the benchmarks during weeks 2-8. The following sections dig deeper into the origin of the skill from each component, considering differences depending on the season and the target WR.

\subsection{Seasonality of the sources of predictability}\label{subsec:seasonal_acc}

\begin{figure*}[h]
 \centerline{\includegraphics[width=33pc]{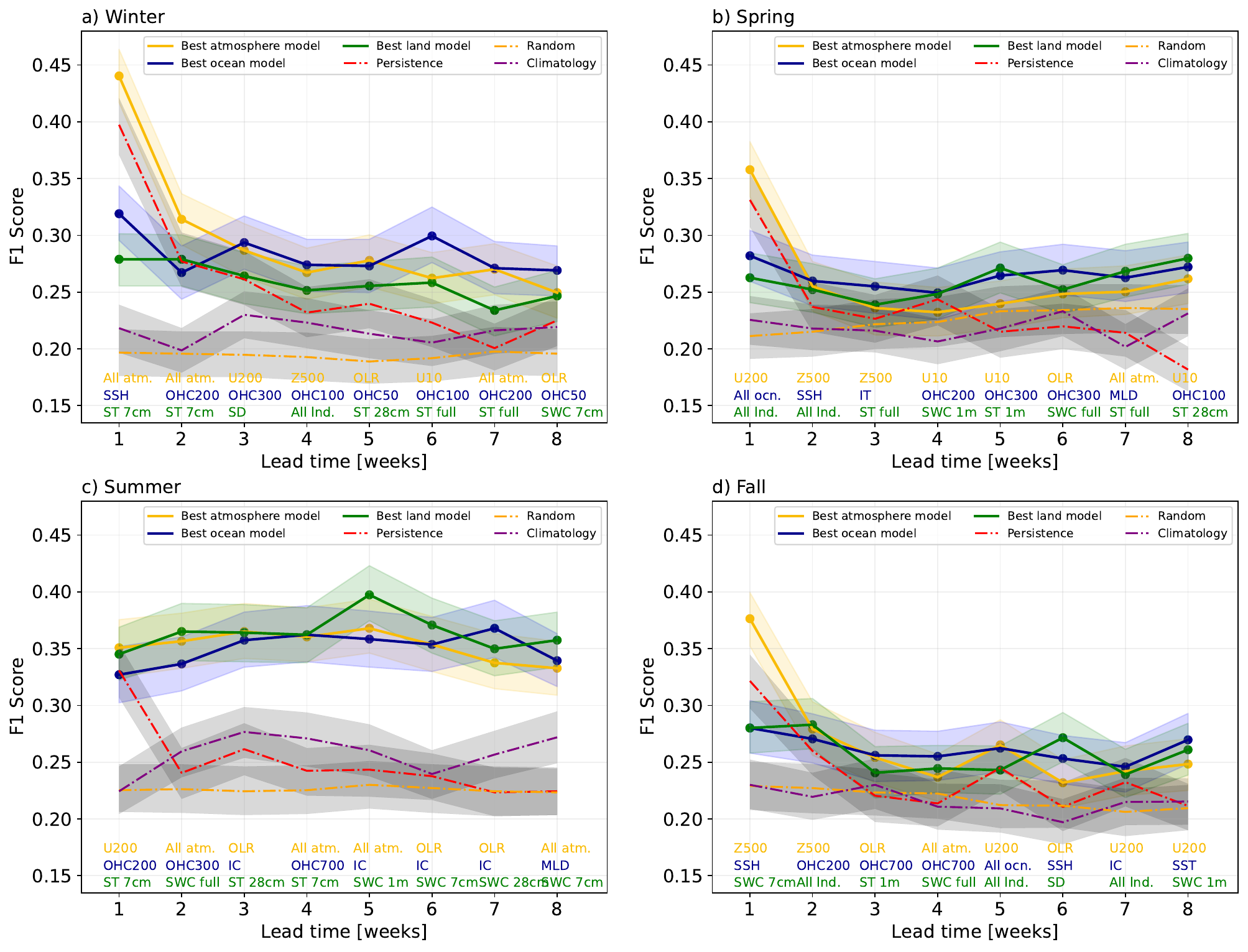}}
  \caption{Same as Figure \ref{f2}d but for S2S forecasts stratified by the season initialization took place.}
  \label{f4}
\end{figure*}

Given the seasonality of the large-scale midlatitude atmosphere and the Earth system in general, certain predictors may provide predictability only during some periods of the year. Some examples include seasonal variations in teleconnections, land heat and moisture fluxes, incident radiation and radiative fluxes, and sea ice thickness and extent. Similar to Figure \ref{f2}d but stratified by season, Figure \ref{f4} shows notable differences in seasonal predictability from the three Earth system components. There is also high variability in the baselines themselves: persistence presents less skill during summer compared to the other seasons, illustrating that baselines can complicate ML model assessments if seasonality is not considered.

When comparing the four seasons in Figure \ref{f4}, it is evident that increased skill from the three components occurs primarily during winter and summer and secondly during spring and fall. These results were somewhat expected because higher baroclinicity and Rossby wave activity during winter yield a more WR-dominated mid-latitude atmosphere, facilitating the understanding of the large scale \citep{kuang2014spatial}. Additionally, teleconnection activity is higher during winter \citep{madden1986seasonal,philander1999review}, providing additional mechanisms for skillful predictions. During summer, there is a predominantly barotropic atmosphere, when the weekly No-WR frequency is higher (it goes from 18\% in the winter to 44\% in the summer). The models do a better job than the baselines distinguishing this pattern, as shown in Figures \ref{f5}e and \ref{f6}s. This seasonality is also reflected in the skill of persistence. During winter, persistence shows increased skill above the other two benchmarks during weeks 1-3, while in summer, this only happens for week 1.

During winter (Figure \ref{f4}a), the atmosphere provides predictability during most lead times, and the ocean-based models show increased skill compared to the benchmarks during weeks 3-8. The atmospheric skill comes from various variables, with the all-atmosphere model performing best on average during weeks 1, 2, and 7. While U200 and Z500 provide meaningful predictability during early lead times, OLR and U10 show increased skill at later lead times (Figure \ref{fa1}a). For the ocean, OHC- and SST-based models offer the highest skill on average after week 3 (Figure \ref{fa1}a). During summer, the predictability coming from the three components is higher than the benchmarks from weeks 2 to 8, and there is a high overlap among the confidence intervals. During the shoulder seasons, the skill of the models relative to the winter and summer decreases notably. However, there are still cases in which there is predictability coming from different components (with moderate uncertainty). The ocean- and land-based models show increased skill during spring in weeks 5-8, with OHC and SWC/ST providing skill often from their respective components. The skill behavior of the three components is similar during fall. Although the skill obtained during fall and spring is lower, individual WRs have sources of predictability within the Earth system that can be further elucidated when stratifying by regime \citep{conil2009contribution,guo2012rebound,thomas2016influence,meehl2021initialized,balmaseda2010impact,petrieetal2015}. These results are discussed in the next Section \ref{sec:results}\ref{subsec:wr_acc}.

\subsection{Variations in predictability sources based on WR} \label{subsec:wr_acc}

\begin{figure*}[h]
 \centerline{\includegraphics[width=33pc]{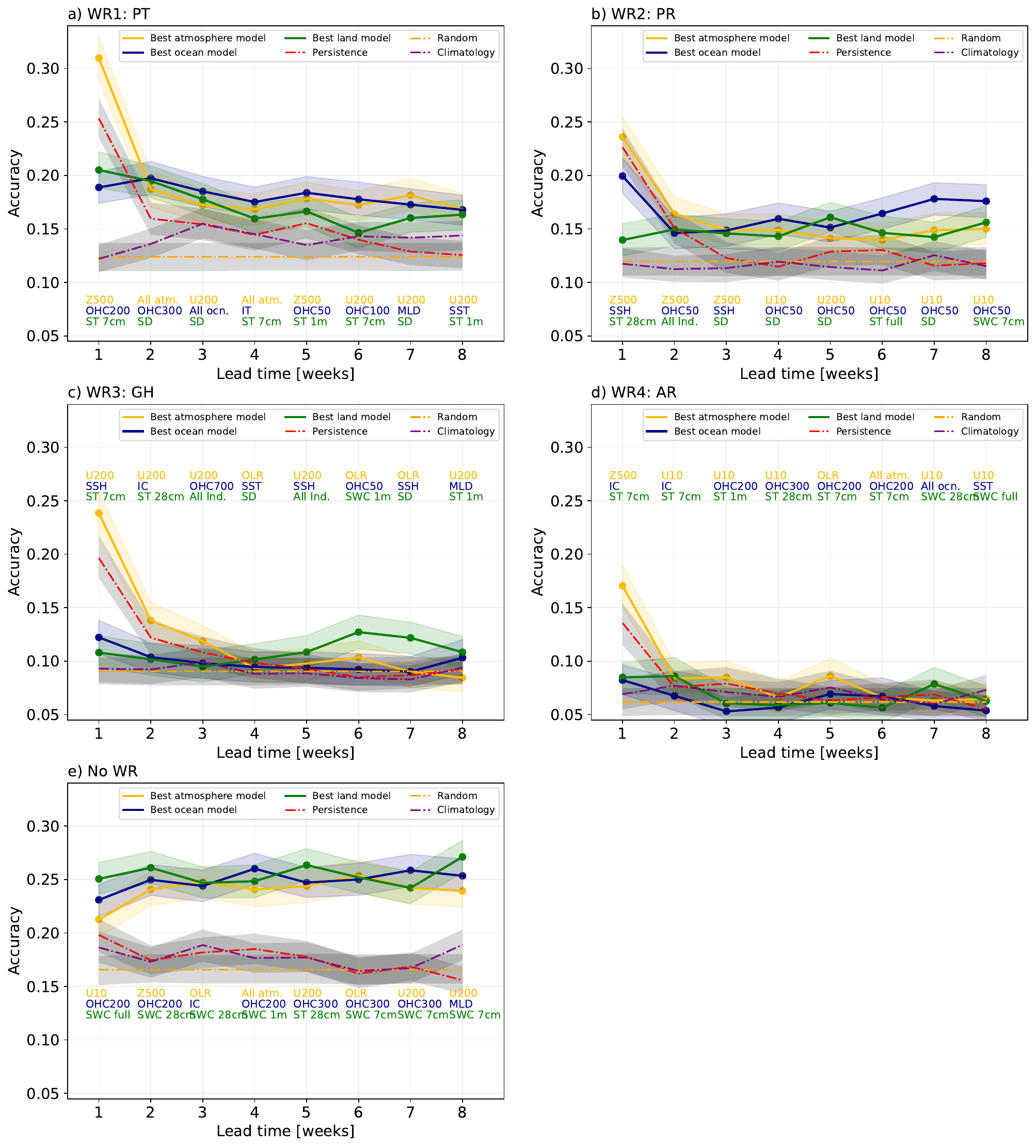}}
  \caption{Same as Figure \ref{f3} but for S2S forecasts stratified by target regime. Accuracy is used instead of the F1 score since there is only one regime (i.e., class) being evaluated. Evaluation is performed on a subset of the test set that contains all samples in which either the ground truth or the prediction (or both) are equal to the regime of interest.}
  \label{f5}
\end{figure*}

\begin{figure*}[h]
 \centerline{\includegraphics[width=39pc]{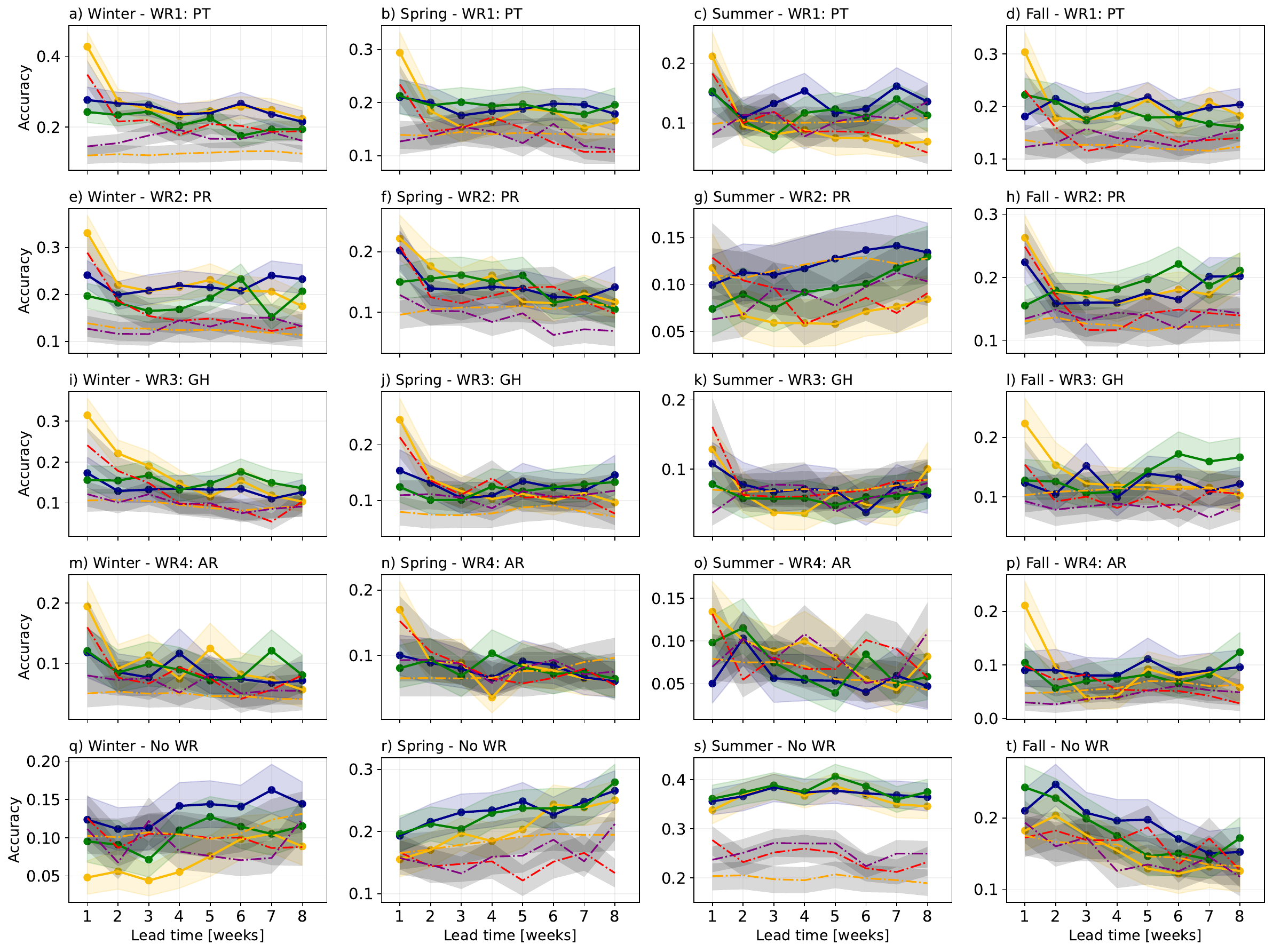}}
  \caption{Same as Figure \ref{f4} stratified by both WR and season. Note that y-axis ranges differ.}
  \label{f6}
\end{figure*}

\begin{figure*}[h]
 \centerline{\includegraphics[width=39pc]{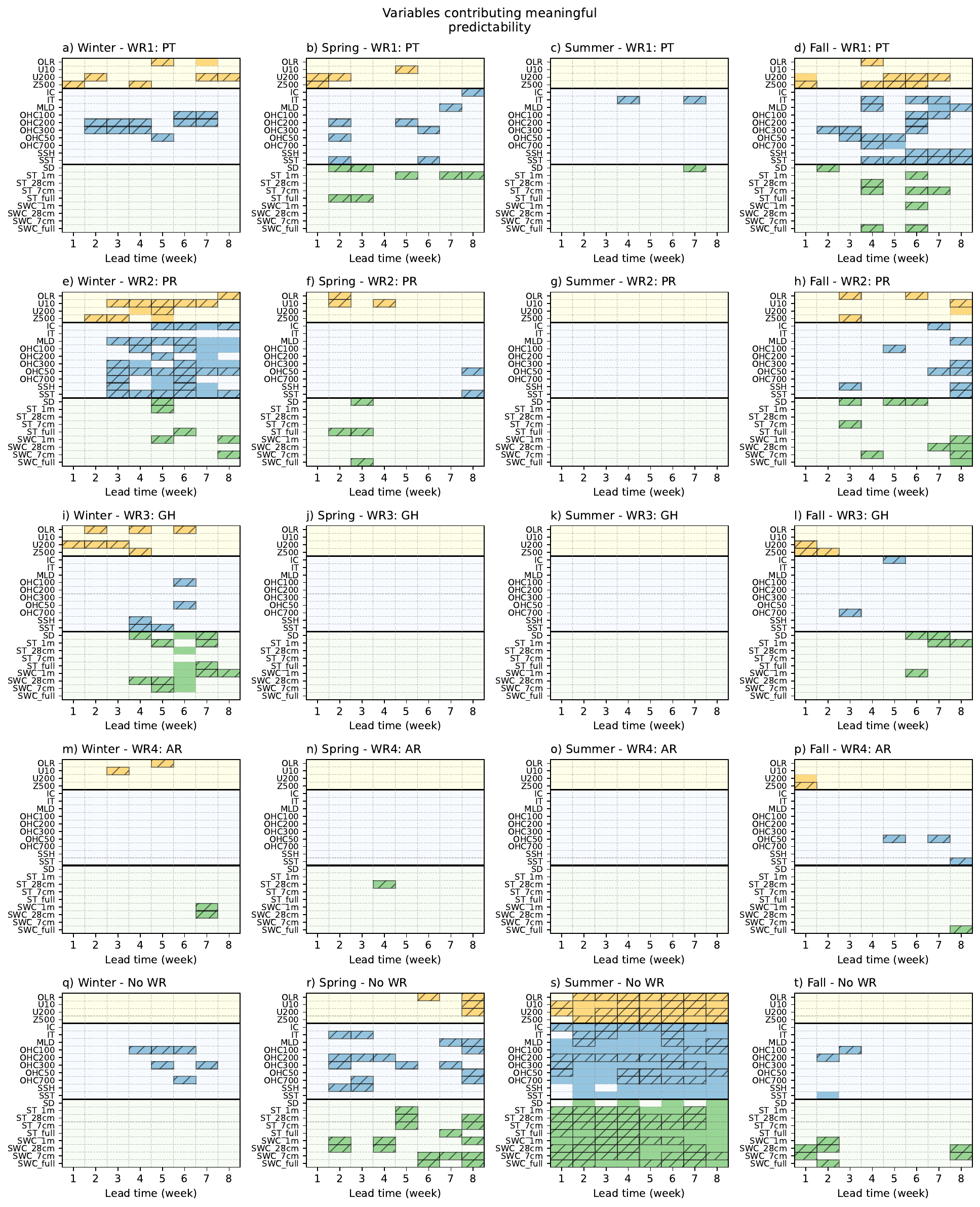}}
  \caption{Same as Figure \ref{f3} stratified by both WR and season.}
  \label{f7}
\end{figure*}

As shown by Figure \ref{f5}, sources of predictability differ for the Pacific Trough, Pacific Ridge, and Greenland High regimes \citep{baldwin2001stratospheric, bueler2021, molina2023subseasonal, hochman2021,kolstad2022}. To further understand the different sources of predictability among regimes, we also explore the seasonal variations in Figure \ref{f6}. Additionally, Figure \ref{f7} shows the variables that contribute meaningful skill for each season and WR. We omitted the Alaskan Ridge regime from this subsection's discussion since prediction skill is not statistically significant or meaningfully different than the considered baselines (Figure \ref{f5}d).

\subsubsection{Pacific Trough}

According to Figure \ref{f5}, the ocean is often (on average) a more substantial contributor to skillful prediction of the Pacific Trough regime from weeks 2 through 8. Additionally, the atmosphere shows skill during most lead times, and the land contributes generally less predictability. According to Figure \ref{f6}a-d, the increased skill is present predominantly during winter, spring, and fall. During fall (Figure \ref{f7}d), the most notable oceanic predictability contributors are SST, SSH, and OHC integrated down to different depths. From the atmosphere and land, U200, Z500, and ST show predictability at several lead times. During winter, the oceanic predictability comes only from OHC, while there are atmospheric predictability contributions from Z500, U200, and OLR (Figure \ref{f7}a). Ocean and land contributions also happen during spring without a clear main contributor (Figure \ref{f7}b), although we note that ST and SD provide skill during some lead times.

\subsubsection{Pacific Ridge}

According to Figure \ref{f5}b, the Pacific Ridge regime shows ocean-based predictability during weeks 3-8, and significant but generally lower predictability during later weeks from the atmosphere and land. When considering different seasons (Figures \ref{f6}e-h and \ref{f7}e-h), there is generally higher atmosphere- and ocean-based predictability during winter. U10 shows relatively high skill during weeks 3-7, which is an indication of stratospheric processes notably affecting the winter occurrence of the Pacific Ridge (Figure \ref{f7}e). From the ocean, there seem to be numerous relevant variables during winter, including IC (weeks 5-8), MLD (weeks 3-8), OHC (weeks 3-8) and SST (weeks 3-8). Predictability contributions are evident during fall, with fewer during spring. During fall (Figure \ref{f7}e), the SD-based models provided skill at several lead times.

\subsubsection{Greenland High}

According to Figure \ref{f5}, the Greenland High regime shows increased atmosphere-based predictability during weeks 1-3 and land during weeks 5-8. Figure \ref{f6}i-l indicates that this predictability behavior occurs primarily in winter and fall. According to Figure \ref{f7}i, the variables that most contribute to winter predictability are U200 (weeks 1-3) and OLR (weeks 2, 4, and 6). Later during the forecast horizon, the skill comes mostly from land: SD, ST, and SWC. The number of models that show meaningful skill during fall decreases notably (Figure \ref{f7}l).

\subsubsection{No-WR class}

According to Figure \ref{f6}q-t, the increased skill in predicting the No-WR class happens primarily in summer, with some variables providing skill in spring, particularly OHC and SWC. However, due to overlapping confidence intervals, it is challenging to identify specific components or predictors as the most relevant during summer. Consequently, no specific variable is highlighted (see Figure \ref{f7}s), and this class is not discussed further.

The results from this section, largely portrayed in Figure \ref{f7}, allow identifying where the skill shown in Figures \ref{f2}d and \ref{f4} comes from. When considering each regime independently, where the trained models exploit predictability from becomes clearer. Oceanic processes seem to be particularly relevant for the Pacific Trough regime during fall and winter. The occurrence of the Pacific Ridge during winter seems to be influenced by a mix of processes from the stratosphere and the ocean. The structure of the upper troposphere and OLR provide predictability during the first few weeks of the forecast for the Greenland High during winter, while the soil water content and temperature have a lagged contribution at lead times greater than 4 weeks. Finally, snow depth seems to contribute some skill to forecasting the Pacific Ridge during autumn and the Greenland High during autumn and winter. In the next section, we further detail where the presented forecasting skill comes from, with an emphasis on the initial states of the Earth system that correspond with accurate forecasts (forecasts of opportunity).

\subsection{Physical processes contributing predictability}\label{subsec:physical}

Recent literature has highlighted the importance of taking advantage of particular initial states of the Earth system that offer increased predictability \citep[namely, forecasts of opportunity;][]{mariotti2020windows,robertson2020toward,mayer2021subseasonal}. In this section, we focus on the initial states in cases when the trained models exhibited significant skill. We do this by compositing standardized anomaly maps of the initial states for the variables and associated correct predictions highlighted in Figure \ref{f7}. Additionally, to identify the regions most relevant to the correct forecasts, we analyze the spatial distribution of average SHapley Additive exPlanations (SHAP) values for accurate predictions \citep{lundberg2017unifiedapproachinterpretingmodel}. SHAP values quantify the contribution of each input feature to the model’s prediction by measuring the change in the expected prediction when conditioning on that feature. Thus, if a feature (or grid cell) consistently exhibits high positive SHAP values during accurate predictions of a specific WR, it indicates that the feature was particularly important for the model’s success.

\subsubsection{Atmospheric precursors}

\begin{figure*}[h]
 \centerline{\includegraphics[width=39pc]{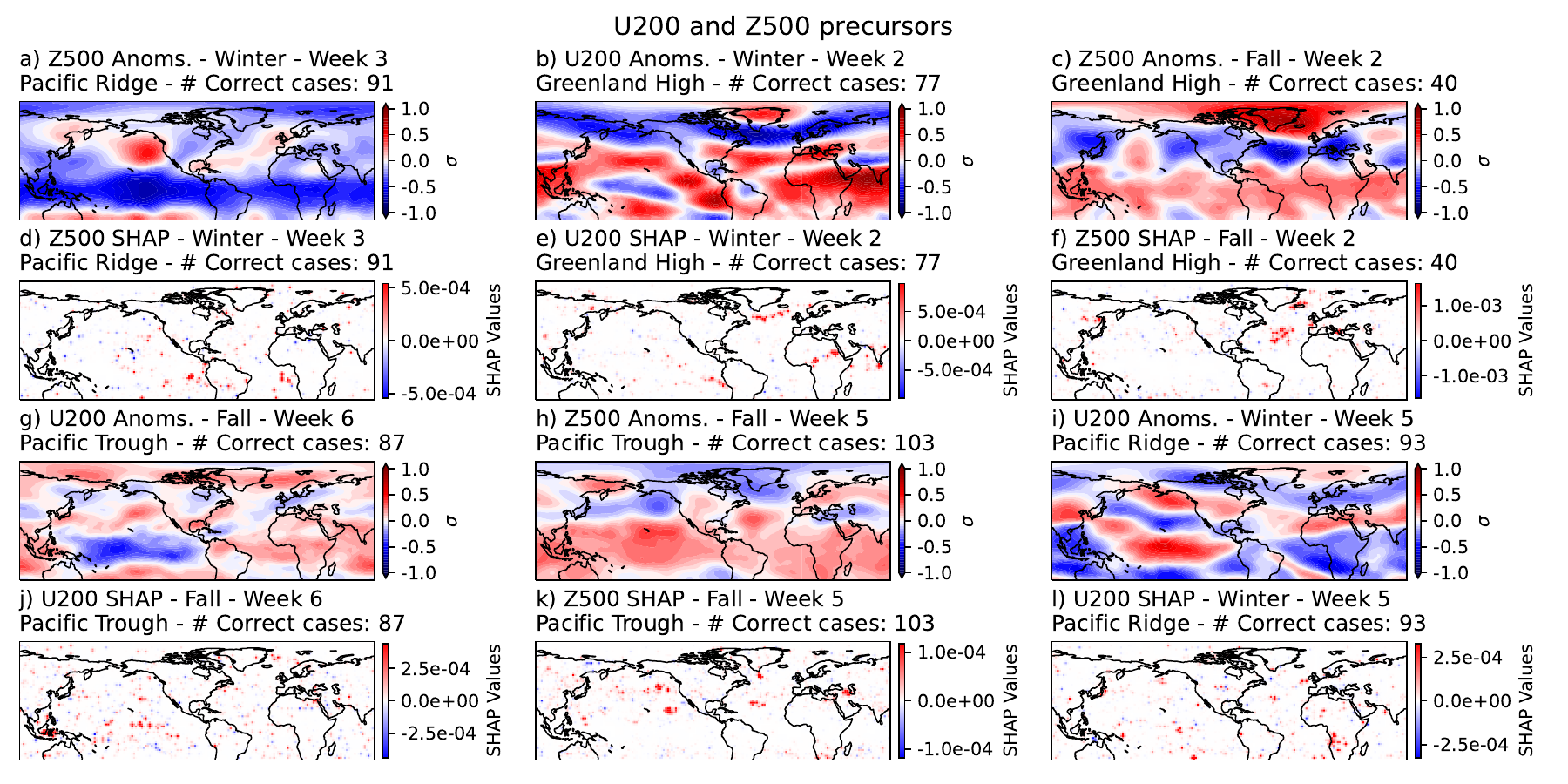}}
  \caption{Initial state standardized anomaly composites of Z500 and U200 corresponding to correct predictions of the specific WR, for the season and lead time indicated in the title of each panel. The average SHAP values corresponding to these accurate predictions are also shown.}
  \label{f8}
\end{figure*}

As shown in Figure \ref{f7}, many atmosphere-based models show skill beyond persistence, especially during winter (Figures \ref{f4}a and \ref{fa1}). U200- and Z500-based models contributed predictability for most WRs during winter, spring, and fall. U200 and Z500 predictability can come from both the persistent anomalies associated with the specific WR and other large-scale features of the Earth system. Figure \ref{f8} shows composited standardized anomalies and SHAP values for when several Z500- and U200-based models achieved accurate predictions for different WRs. During early lead times (weeks 2-3), the Pacific Ridge (winter) and Greenland High (winter and fall) regimes show increased predictability from these two variables (Figure \ref{f7}e,i,l). Correct Pacific Ridge predictions during week 3 were associated with negative Z500 anomalies over the tropics, with positive anomalies over the northeastern Pacific Ocean (Figure \ref{f8}a). Clusters of positive and high SHAP values indicate that the tropics contributed more to the accurate predictions relative to the persistent circulation over the northeastern Pacific (Figure \ref{f8}d). Similar anomalies occur when accurately predicting week 2, but SHAP values are higher over the location of the ridge itself (not shown), indicating a higher contribution from persistence at a shorter lead time. 

For the Greenland High regime, U200 showed skill during weeks 1-3 in winter (Figure \ref{f7}i). According to Figure \ref{f8}b,e, predictability contributions come from both the Greenland High persistent circulation (negative U200 anomalies south of Greenland) and increased eastward wind over the eastern tropical Pacific and northern Indian Ocean. For week 3, the SHAP spatial pattern is similar but with higher values over the tropics (not shown). For the Greenland High during fall, only the persistent large-scale extra-tropical structure seems to provide predictability out to week 2 (Figures \ref{f7}l and \ref{f8}c,f). For longer lead times in the fall, Z500 and U200 yield skill in predicting the Pacific Trough and Pacific Ridge regimes. Westward anomalies of U200 over the tropical Pacific Ocean and increased Z500 over the tropics are associated with accurate predictions of the Pacific Trough regime during weeks 4-7 (Figures \ref{f7}d and \ref{f8}g,h,j,k). The opposite is true for accurate predictions of the Pacific Ridge during winter (Figure \ref{f8}i,l), which were associated with positive eastward upper-level flow over the tropical Pacific and upper-level divergence over the maritime continent. These large-scale characteristics related to Pacific Trough and Pacific Ridge occurrence resemble the Walker circulation response for the two opposite states of the ENSO (El Niño and La Niña, respectively).

\begin{figure*}[h]
 \centerline{\includegraphics[width=39pc]{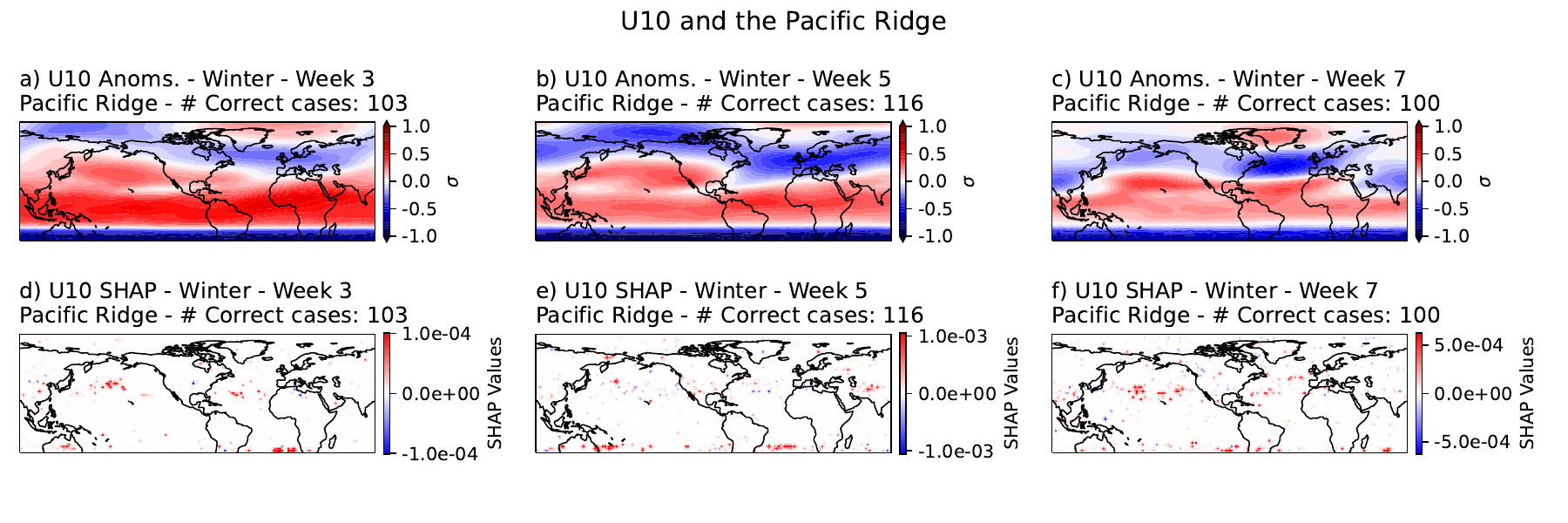}}
  \caption{Initial state standardized anomaly composites of U10 corresponding to correct predictions of the specific WR, for the season and lead time indicated in the title of each panel. The average SHAP values corresponding to these accurate predictions are also shown.}
  \label{f9}
\end{figure*}

U10 in Figure \ref{f7}e consistently shows skill for predicting the Pacific Ridge during winter. Predictability from U10 at weeks 3-7 is associated with the presence of upper-level easterly anomalies over the Southern Hemisphere tropics (blue in Figure \ref{f9}a,b,c), which was also relevant for the XGBoost models yielding a correct Pacific Ridge prediction (Figure \ref{f9}d,e,f). Accompanying this signal are westerly U10 anomalies along the equator and North Pacific Ocean (from around -5$^\circ$N to 35$^\circ$N). Finally, the composites show easterly anomalies near the North Pole west of North America and around 60$^\circ$N east of North America, which is potentially related to structural changes of the stratospheric polar vortex. Although the stratospheric winds both at the tropics and near the pole show relevance for S2S prediction of the Pacific Ridge, it is difficult to link its occurrence directly with SSW events or specific phases of the QBO. On the one hand, the slowing of the stratospheric polar vortex is not zonally uniform. On the other hand, the composite U10 anomalies within the tropics are not strong at the equator \citep[where the QBO is classically defined;][]{baldwin2001quasi} but rather around 20$^\circ$S-30$^\circ$S. However, the robustness of the relationship (see Figure \ref{f7}e) justifies future more in-depth research on this predictor.

\begin{figure*}[h]
 \centerline{\includegraphics[width=39pc]{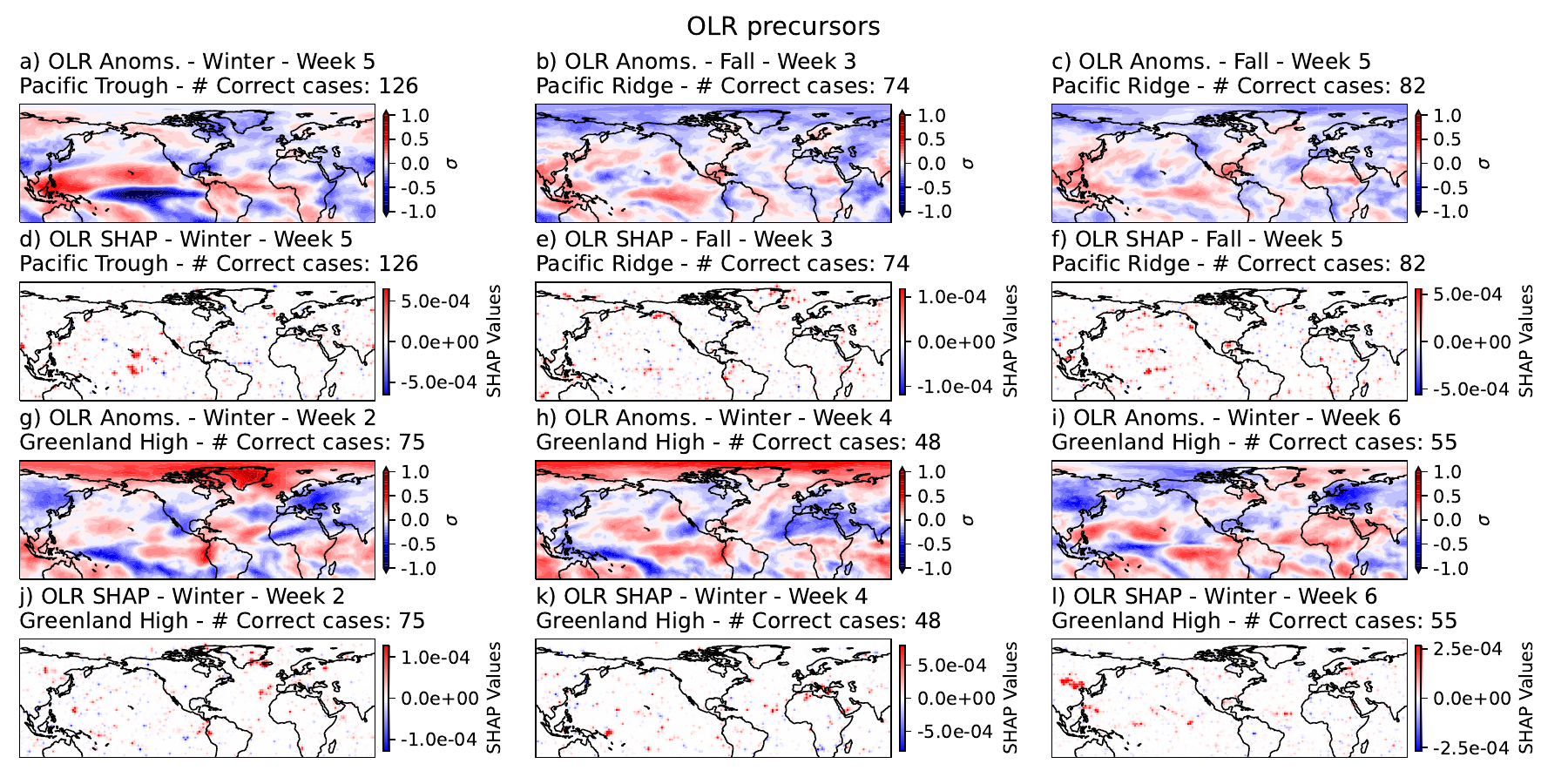}}
  \caption{Initial state standardized anomaly composites of OLR corresponding to correct predictions of the specific WR, for the season and lead time indicated in the title of each panel. The average SHAP values corresponding to these accurate predictions are also shown.}
  \label{f10}
\end{figure*}

According to Figure \ref{f7}a,h,i, OLR provides predictability during some lead times for the Pacific Trough and Greenland High regimes in winter, and for the Pacific Ridge in fall. The composite anomalies and SHAP fields show that accurate prediction of the Pacific Trough relates to increased convection over the central tropical Pacific (El-Niño-like signal; Figure \ref{f10}). In contrast, the OLR pattern is weak for the Pacific Ridge during fall. The OLR pattern is mixed for the Greenland High in winter; decreased cloudiness over Greenland (persistence) plays a relevant role for week 2 (Figure \ref{f10}g,j), while enhanced cloudiness over east Asia and central Europe provides increased predictability for week 6. When OLR provides meaningful predictive skill for the Greenland High regime, the tropics show a pattern similar to phases 6 and 7 of the MJO \citep{wang2018mjo}, with enhanced convection east of the Maritime continent. These particular MJO phases preceding the onset of Greenland Blocking are consistent with \cite{parker2018ensemble}, who showed how the progression of the MJO was related to a negative NAO (and also the onset of a Greenland High regime).

The results associated with U10 and OLR as precursors highlight the complexity of the interactions between the tropical troposphere and the stratospheric vortex that influence the occurrence of weather regimes \citep{lee2019wintertime,barnes2019tropospheric,green2019evaluating}, suggesting that the stratosphere may be more important for the onset of the Pacific Ridge, while tropical OLR may be closer tied to the Greenland High and Pacific Trough regimes.

% \newpage
\subsubsection{Oceanic precursors}

\begin{figure*}[h]
 \centerline{\includegraphics[width=39pc]{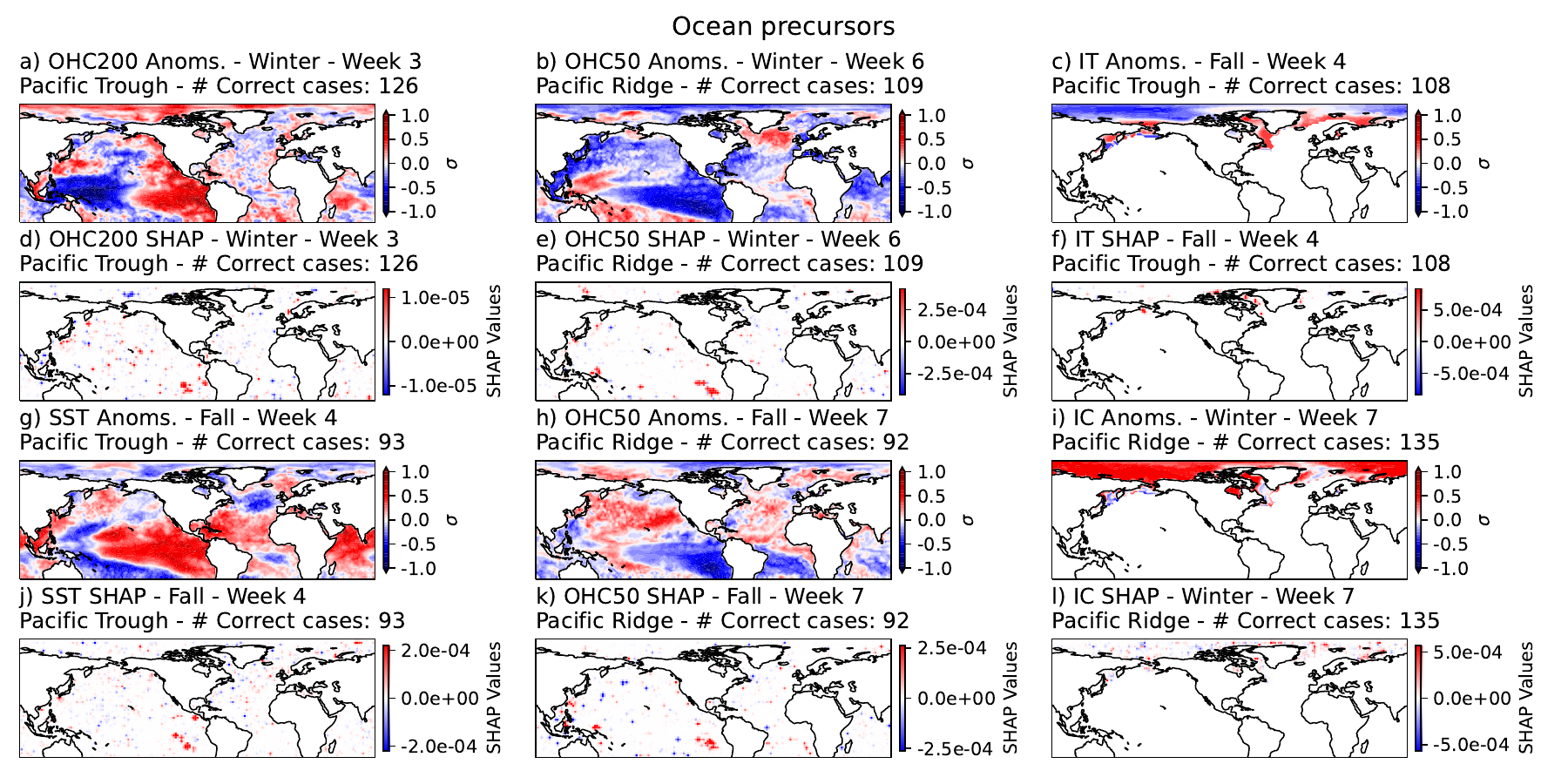}}
  \caption{Initial state standardized anomaly composites of OHC, SST, IT, and IC corresponding to correct predictions of the specific WR, for the season and lead time indicated in the title of each panel. The average SHAP values corresponding to these accurate predictions are also shown.}
  \label{f11}
\end{figure*}

The ocean was a significant source of predictability for the Pacific Trough and Pacific Ridge regimes during winter and fall (Figure \ref{f7}a,d,e,h). Figure \ref{f11}a,b,g,h shows composites of OHC and SST standardized anomalies for accurate predictions of the Pacific Trough and Pacific Ridge regimes. Although other ocean variables also contributed significant predictability (e.g., SSH, MLD, and OHC at different depths; Figure \ref{f7}), the associated spatial patterns were similar. As expected \citep{lhereux2006observed,furtado2012linkages,deser2017thenorthern}, the tropical Pacific Ocean and ENSO play a relevant role (Figure \ref{f11}a,b,g,h). Consistent with \cite{straus2007}, \cite{vigaud2018predictability}, and \cite{molina2023subseasonal}, an El-Niño-like signal during fall and winter helps accurately predict the Pacific Trough regime. Analogously, a La-Niña-like tropical Pacific provides skill for predicting the Pacific Ridge regime. Although ocean-based predictability and ENSO are more consistent sources of predictability for the Pacific Trough regime during fall and the Pacific Ridge during winter, the spatial distribution of anomalies and SHAP are similar (Figure \ref{f11}a-b,d-e,g-h,j-k). It is worth noting that ENSO-derived predictability is not only identifiable in the ocean since atmosphere-based models can also accurately predict the Pacific Trough and Pacific Ridge regimes using atmospheric patterns related to ENSO, such as those shown in Figures \ref{f8}a,g,h,i and \ref{f10}a.

IT provides predictability for the Pacific Trough during fall and IC for the Pacific Ridge during winter (Figure \ref{f7}d,f). Accurate Pacific Trough predictions in fall are associated with increased IT over the Labrador Sea, Baffin Bay, Bering Sea, and Chukchi Sea, with negative anomalies over the North Pole (week 4 is shown in Figure \ref{f11}c,f, but weeks 6 and 7 are similar). Increased IC over the whole polar region helps predict the Pacific Ridge in winter (week 7 is shown in Figure \ref{f11}i,l, but similar results for weeks 5-8). These results for the Pacific Trough and Pacific Ridge regimes are in agreement with \cite{gervais2024}, indicating that less ice over the North Pole favors the occurrence of Aleutian lows.

% \newpage
\subsubsection{Land precursors}

\begin{figure*}[h]
 \centerline{\includegraphics[width=39pc]{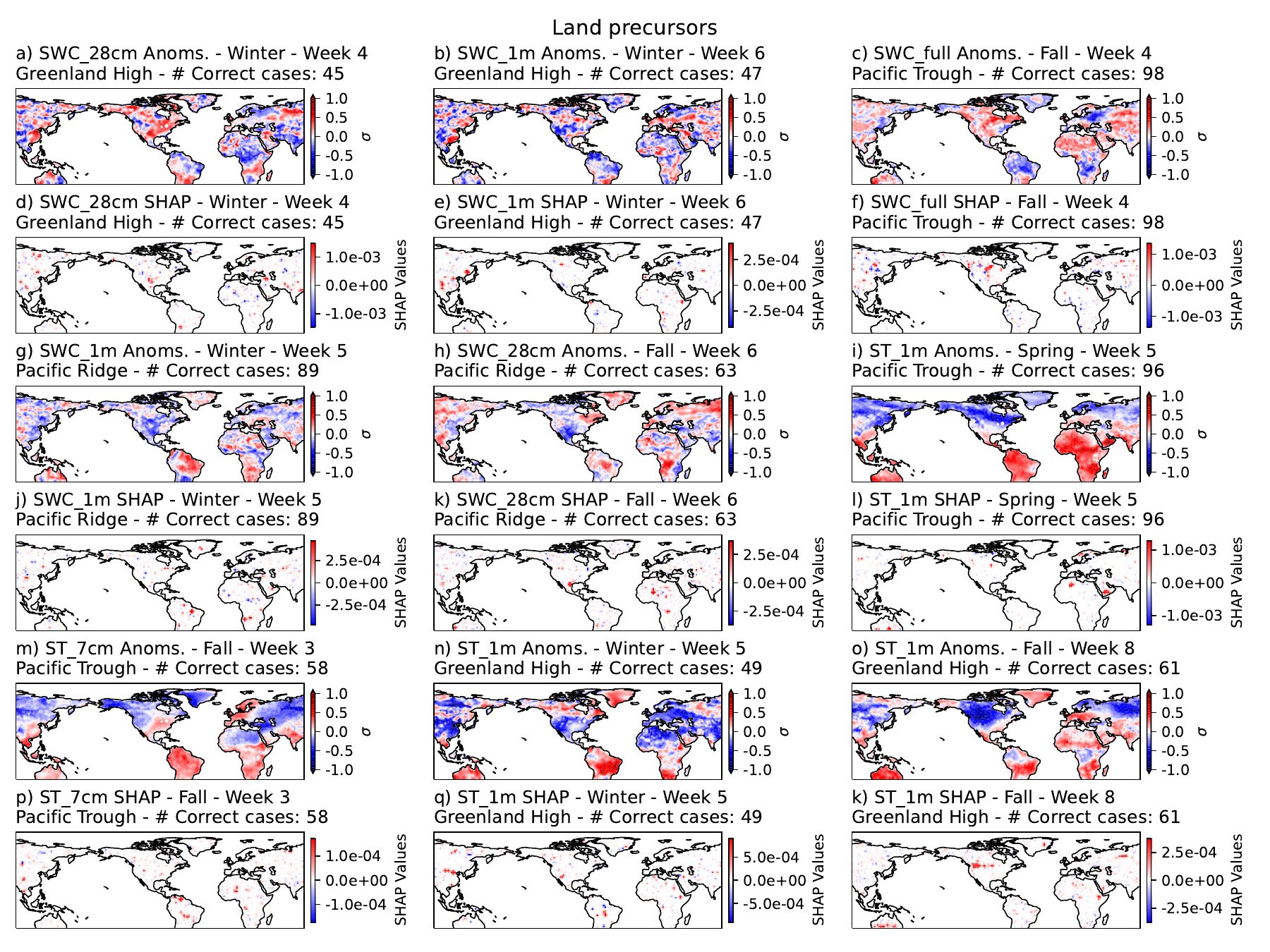}}
  \caption{Initial state standardized anomaly composites of SWC and ST corresponding to correct predictions of the specific WR, for the season and lead time indicated in the title of each panel. The average SHAP values corresponding to these accurate predictions are also shown.}
  \label{f12}
\end{figure*}

Figure \ref{f7} shows that different land variables provide meaningful predictability during different lead times and seasons for the Pacific Trough, Pacific Ridge, and Greenland High regimes. SWC provides increased skill at several lead times when predicting the Pacific Trough during fall (Figure \ref{f7}d), the Pacific Ridge during winter and spring (Figure \ref{f7}e,f), and the Greenland High during winter (Figure \ref{f7}i). For the Greenland High in winter, models based on SWC averaged down to 28cm and 1m provide predictive skill during weeks 4-6 and 6-8, respectively. Given how sparse and noisy the SWC composites and corresponding SHAP values are (see Figure \ref{f12}a,b,d,e), it is difficult to highlight specific regions as relevant. However, both layers and the various lead times have positive SWC anomalies over southeastern Asia, with high but scattered SHAP values over that region (Figure \ref{f12}a,b,d,e). The anomalies and SHAP fields are more consistent across lead times for SWC-1m than for SWC-28cm (not shown). For the Pacific Ridge during winter and fall, negative anomalies of SWC over the western U.S. and positive anomalies over the Amazon rainforest are notable (Figure \ref{f12}g,h,j,k). However, SHAP values don't overlap over those regions making it difficult to draw conclusions (Figure \ref{f12}j,k). The SWC-full anomalies associated with increased predictability for the Pacific Trough in fall seem to be associated with increased SWC over North America, but the SHAP fields are sparse, similar to the Greenland High and Pacific Ridge results. Although earlier results in Section \ref{sec:results}\ref{subsec:wr_acc} indicate that there is predictive skill from SWC, further exploration of these potential links is needed because our approach offers limited interpretation.

ST contributes to skillful prediction of the Pacific Trough regime during spring and fall and Greenland High during winter and fall (see Figure \ref{f7}b,d,i,l). Pacific Trough's increased predictability is associated with warmer soil temperatures over tropical South America and Africa and cooler temperatures over most of the extratropics, mainly eastern Europe, northern Asia, and Canada (see Figure \ref{f12}i,l,m,p). This land temperature pattern, given its global extent, could be a footprint of the El Niño phase of ENSO. ST-1m-based models are skillful when predicting the Greenland High during weeks 5 and 7 during winter and weeks 7 and 8 during fall (Figure \ref{f7}i,l). The composites in these cases show negative anomalies over western North America, southeastern Asia, and western Russia, with positive anomalies over the Amazon in South America. The pattern shown for ST in Figure \ref{f12}n slightly resembles the one shown in Figure \ref{f10}h,i, indicating that the predictability from OLR and ST may have the same process as its origin, such as the MJO-extratropic-tropospheric pathways described by \cite{lee2019wintertime}, \cite{green2019evaluating}, and \cite{barnes2019tropospheric}.

\begin{figure*}[h]
 \centerline{\includegraphics[width=39pc]{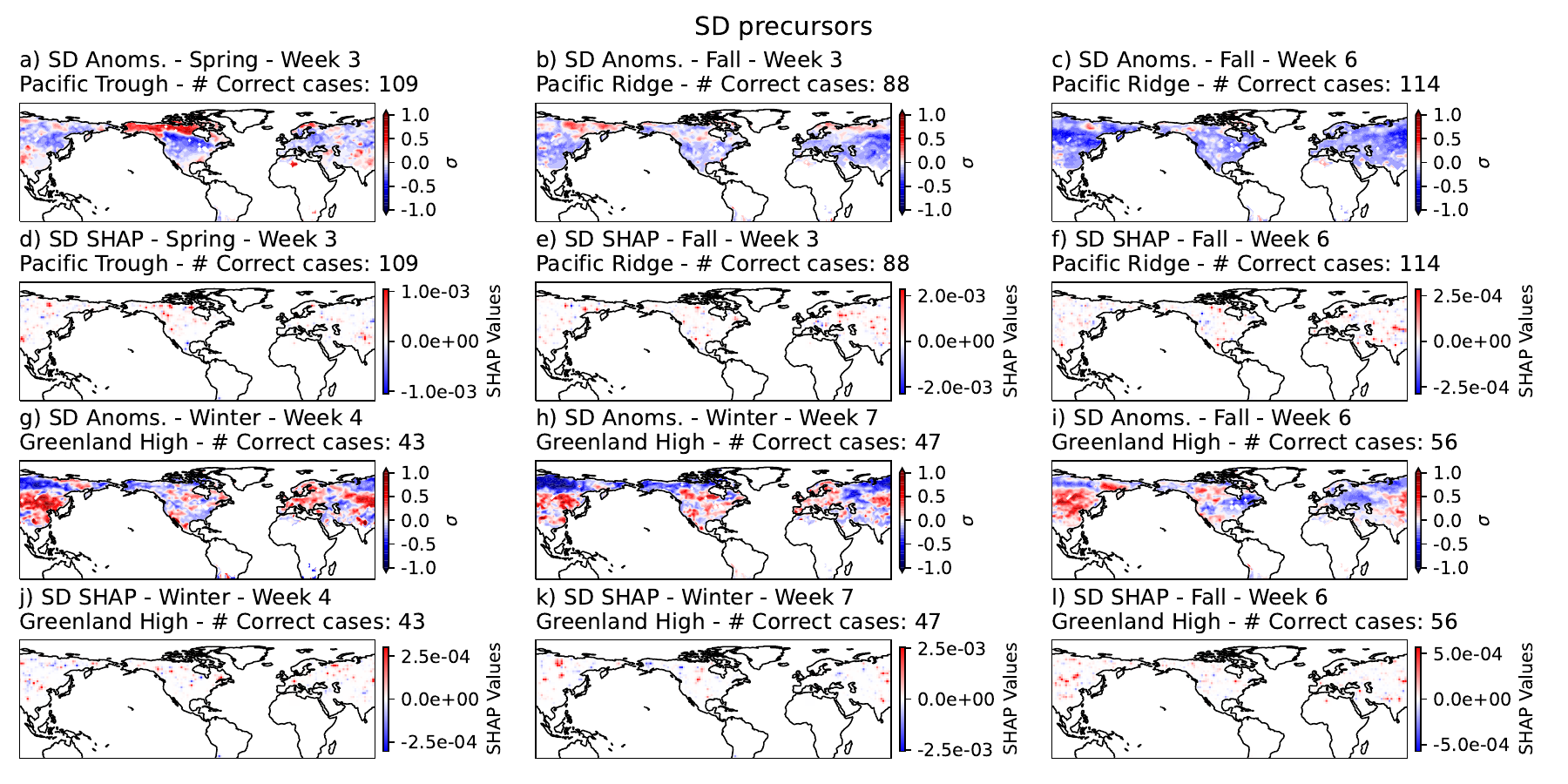}}
  \caption{Initial state standardized anomaly composites of SD corresponding to correct predictions of the specific WR, for the season and lead time indicated in the title of each panel. The average SHAP values corresponding to these accurate predictions are also shown.}
  \label{f13}
\end{figure*}

Another predictability contributor for North American WRs is snow depth (SD). According to Figure \ref{f7}b,h,i,l, SD-based models provided skill for the Pacific Trough in spring (weeks 2 and 3), the Pacific Ridge in fall (weeks 3, 5, and 6), and the Greenland High in winter (weeks 4, 6, and 7) and fall (weeks 6 and 7). The increased predictability of the Pacific Trough during spring is associated with increased SD over Alaska and Northern Canada and decreased SD over the U.S. Pacific Northwest (Figure \ref{f13}a,d). For the Pacific Ridge in fall, the increased predictive skill is derived from negative SD anomalies over most of the Northern extra-tropics, mainly over Asia (Figures \ref{f13}b,c,e,f). For the Greenland High, increased predictability is associated with increased SD over Asia, and decreased SD over northern Russia (Figure \ref{f13}g,h,i,j,k,l). These results support studies that have shown a significant lagged influence of snowpack over the mid-latitude large-scale atmosphere \citep{peings2011snow,orsolini2013impact,thomas2016influence}. However, further research on the mechanisms associated with this potential source of predictability is needed.

\section{Conclusions}\label{sec:conclusions}

In this study, we used ML to learn more about the sources of predictability of North American WRs at S2S lead times. We trained several XGBoost models with an input variable from one of three Earth system components to assess their relative importance at various lead times and seasons. Similar to the results from \cite{richter2024quantifying}, we found that atmospheric variables were better predictors during weeks 1-2 of the forecast, with the ocean providing only slightly higher predictability than the other two components on average after week 3. In general, the ML models that included all variables from the land or ocean did not necessarily present the best skill, possibly due to competing predictive signals or limitations with using correlated variables. Thus, while convenient, it is not advisable to train an ML model that includes all predictors without special treatment or weighting. Contrary to our expectations, fewer input features could provide more predictability when using ML, particularly for slower-evolving land and ocean processes.

Our results also show how highly seasonally-dependent the sources of predictability are. We demonstrate that the S2S weather regime predictability increases during winter compared to spring and fall. For winter, the ocean and atmosphere provide more predictability than the land, while during spring, the ocean- and land-based models show higher skill than the atmosphere-based models. When initialized in summer, the models have virtually no skill in predicting the specific weather regime; skill instead came from the models accurately forecasting the No-WR class. The forecast skill and the sources of predictability were also dependent on the regime being predicted. The models showed higher skill when predicting the Pacific Trough and Pacific Ridge regimes compared to the Greenland High and Alaskan Ridge. When stratifying by WR and season in Section \ref{sec:results}\ref{subsec:wr_acc}, results indicate that the ocean provides predictability for the Pacific Trough and Pacific Ridge regimes during fall and winter. Stratospheric processes are especially relevant for the Pacific Ridge during winter. OLR-related signals help forecast the Greenland High during some weeks in winter. The structure of the upper troposphere corresponding both to persistence and other large-scale features was relevant for the Pacific Trough, Pacific Ridge, and Greenland High regimes during early lead times. Additionally, some land-based models show significant skill during winter, spring, and fall.

In Section \ref{sec:results}\ref{subsec:physical}, we provide a deeper analysis of the initial conditions associated with accurate forecasts and local feature contributions using the SHAP method. The results indicate that persistent tropospheric winds and convection, at times associated with weather regimes, were useful during early lead times (weeks 2-3), while the Walker circulation and tropical and subtropical convection potentially associated with ENSO and MJO were relevant features during later lead times. A spatial pattern consisting of negative stratospheric wind anomalies over the southern tropics (south of 15$^\circ$S) shows a robust relationship with the occurrence and predictability of the Pacific Ridge. Although this pattern may be related to a phase of the QBO and/or sudden stratospheric warming events, future research is needed to confirm their influence. From the ocean, most of the predictability is associated with distinct phases of the ENSO. El Niño provides skill for the Pacific Trough and La Niña for the Pacific Ridge. Additionally, ice thickness and concentration provided skill for predicting the Pacific Trough during fall and the Pacific Ridge during winter, respectively. From land, the spatial patterns of SWC (soil water content) associated with correct predictions are sparse and inconclusive; thus, further research is needed to better understand the influence of soil moisture as a source of predictability for weather regimes. ST (soil temperature) provides some predictability, potentially acting as a footprint of ENSO. SD (snow depth) also shows predictive skill for the Pacific Trough and Pacific Ridge regimes, but more research on the physical mechanisms is needed. ENSO's footprint was evident in the three Earth system components, and when comparing MJO and ENSO contributions, our results agree with \cite{mayer2024} that ENSO is a more meaningful contributor to S2S predictability.

The sources of predictability identified in this study are relevant beyond large-scale circulation patterns. The sources of predictability also present potential for improving the forecasting skill of surface weather, given how closely related regimes are to temperature and precipitation anomalies (see Figure \ref{f1}), storm tracks, and extreme events \citep{lee2023new,jennrich2024}. Future work should focus on more deeply understanding these sources of predictability, including testing causal hypotheses and quantifying how they may change in a warming climate. This study could also be expanded to incorporate other Earth system components and processes, delve further into the predictors herein, or include data from Earth system numerical models. Our study is subject to limitations associated with clustering methods, wherein the regimes may be sensitive to the length of the observational record; a limitation that future work should also explore. Combined with domain knowledge, ML continues to present opportunities to elucidate sources of predictability.

%%%%%%%%%%%%%%%%%%%%%%%%%%%%%%%%%%%%%%%%%%%%%%%%%%%%%%%%%%%%%%%%%%%%%
% ACKNOWLEDGMENTS
%%%%%%%%%%%%%%%%%%%%%%%%%%%%%%%%%%%%%%%%%%%%%%%%%%%%%%%%%%%%%%%%%%%%%
\acknowledgments
JSPC and MJM were supported by a University of Maryland Grand Challenges Seed Grant. MJM was also supported by the National Science Foundation (NSF) under Grant Number 2425735. Computing and data storage resources were provided by the Computational and Information Systems Laboratory at the NSF National Center for Atmospheric Research (NCAR), which is a major facility sponsored by the NSF under Cooperative Agreement No. 1852977. The authors acknowledge the use of OpenAI's ChatGPT for assistance in developing and refining the code used for analysis and visualization. The authors also thank the three anonymous reviewers for their valuable insights, which increased the quality of the manuscript.

\datastatement
ERA5 and SODA reanalyses can be obtained from the NSF NCAR Research Data Archive (https://doi.org/10.5065/D6X34W69 and https://doi.org/10.5065/HBTB-R521). Software developed for this study is available as open source at the GitHub repository: \url{https://github.com/jhayron-perez/WR_Predictability}.

\newpage

\appendix[A] 

\appendixtitle{Variables that contributed meaningful predictability depending on the season and weather regime, separately}

% \subsection*{Appendix section}

\begin{figure}[h]
\center
\includegraphics[width=19pc]{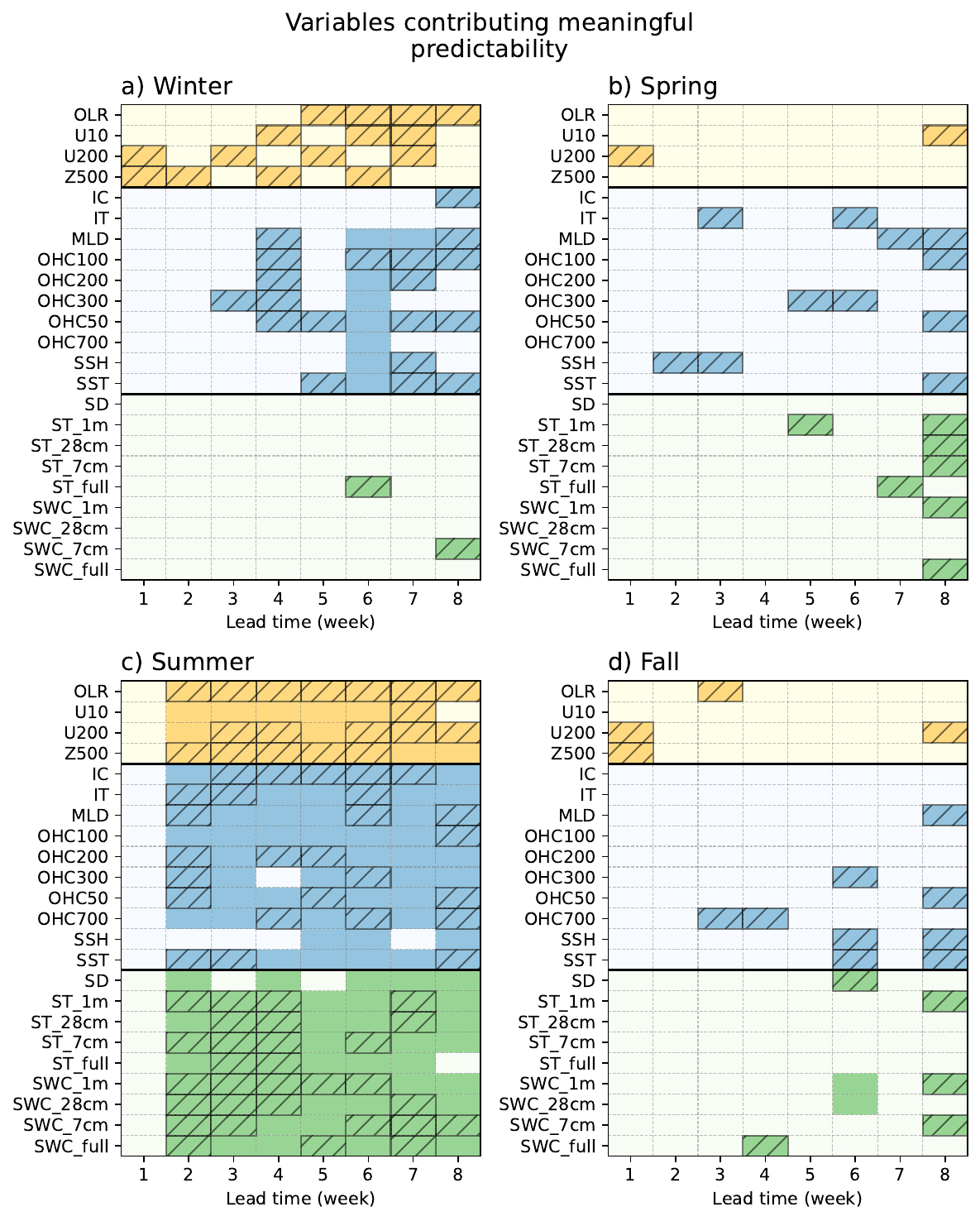}
\caption{Same as Figure \ref{f3} but stratified by the season in which the forecasts were initialized.}\label{fa1}
\end{figure}

\begin{figure}[h]
\center
\includegraphics[width=19pc]{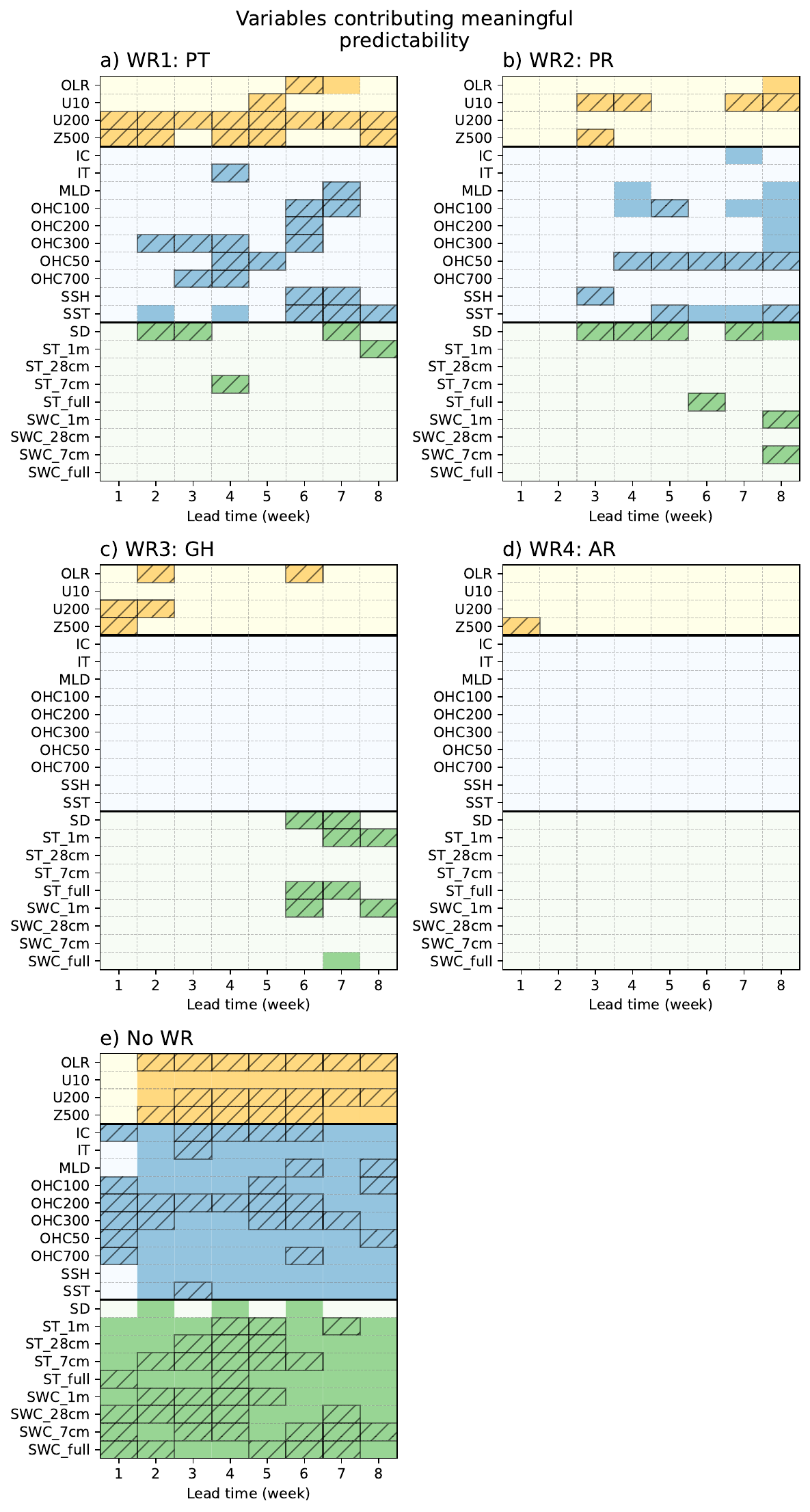}
\caption{Same as Figure \ref{f3} but stratified by target weather regime.}\label{fa2}
\end{figure}

Figures \ref{fa1} and \ref{fa2} show the variables contributing meaningful and high predictability during different seasons and when aiming to predict different weather regimes. These are shown for completeness, but a more detailed version is shown in Figure \ref{f7}.

\newpage

\bibliographystyle{ametsocV6}

\begin{thebibliography}{95}
\providecommand{\natexlab}[1]{#1}
\providecommand{\url}[1]{\texttt{#1}}
\renewcommand{\UrlFont}{\rmfamily}
\providecommand{\urlprefix}{URL }
\expandafter\ifx\csname urlstyle\endcsname\relax
  \providecommand{\doi}[1]{https://doi.org/\discretionary{}{}{}#1}\else
  \providecommand{\doi}{https://doi.org/\discretionary{}{}{}\begingroup \urlstyle{rm}\Url}\fi
\providecommand{\eprint}[2][]{\url{#2}}


\bibitem[{Alexander(1992)}]{alexander1992midlatitude}
Alexander, M.~A., 1992: {Midlatitude atmosphere--ocean interaction during El Ni{\~n}o. Part I: the North Pacific Ocean}. \textit{J.\ Climate}, \textbf{5~(9)}, 944--958, \doi{10.1175/1520-0442(1992)005<0944:MAIDEN>2.0.CO;2}.

\bibitem[{Baldwin et~al.(2001)}]{baldwin2001quasi}
Baldwin, M., and Coauthors, 2001: The quasi-biennial oscillation. \textit{Rev. Geophys.}, \textbf{39~(2)}, 179--229, \doi{10.1029/1999RG000073}.

\bibitem[{Baldwin and Dunkerton(2001)Baldwin, and Dunkerton}]{baldwin2001stratospheric}
Baldwin, M.~P., and T.~J. Dunkerton, 2001: Stratospheric harbingers of anomalous weather regimes. \textit{Science}, \textbf{294~(5542)}, 581--584, \doi{10.1126/science.1063315}.

\bibitem[{Baldwin et~al.(2003)Baldwin, Stephenson, Thompson, Dunkerton, Charlton,, and O'Neill}]{baldwin2003stratospheric}
Baldwin, M.~P., D.~B. Stephenson, D.~W. Thompson, T.~J. Dunkerton, A.~J. Charlton, and A.~O'Neill, 2003: Stratospheric memory and skill of extended-range weather forecasts. \textit{Science}, \textbf{301~(5633)}, 636--640, \doi{10.1126/science.1087143}.

\bibitem[{Balmaseda et~al.(2010)Balmaseda, Ferranti, Molteni,, and Palmer}]{balmaseda2010impact}
Balmaseda, M.~A., L.~Ferranti, F.~Molteni, and T.~N. Palmer, 2010: {Impact of 2007 and 2008 Arctic ice anomalies on the atmospheric circulation: Implications for long-range predictions}. \textit{Quart.\ J.\ Roy.\ Meteor.\ Soc.}, \textbf{136~(652)}, 1655--1664, \doi{10.1002/qj.661}.

\bibitem[{Balsamo et~al.(2009)Balsamo, Beljaars, Scipal, Viterbo, van~den Hurk, Hirschi,, and Betts}]{balsamo2009revised}
Balsamo, G., A.~Beljaars, K.~Scipal, P.~Viterbo, B.~van~den Hurk, M.~Hirschi, and A.~K. Betts, 2009: {A revised hydrology for the ECMWF model: Verification from field site to terrestrial water storage and impact in the Integrated Forecast System}. \textit{Journal of hydrometeorology}, \textbf{10~(3)}, 623--643, \doi{10.1175/2008JHM1068.1}.

\bibitem[{Barnes et~al.(2019)Barnes, Samarasinghe, Ebert-Uphoff,, and Furtado}]{barnes2019tropospheric}
Barnes, E.~A., S.~M. Samarasinghe, I.~Ebert-Uphoff, and J.~C. Furtado, 2019: Tropospheric and stratospheric causal pathways between the mjo and nao. \textit{Journal of Geophysical Research: Atmospheres}, \textbf{124~(16)}, 9356--9371, \doi{10.1029/2019JD031024}, \urlprefix\url{https://agupubs.onlinelibrary.wiley.com/doi/abs/10.1029/2019JD031024}, \eprint{https://agupubs.onlinelibrary.wiley.com/doi/pdf/10.1029/2019JD031024}.

\bibitem[{Becker et~al.(2022)Becker, Kirtman, L’Heureux, Mu{\~n}oz,, and Pegion}]{becker2022decade}
Becker, E.~J., B.~P. Kirtman, M.~L’Heureux, {\'A}.~G. Mu{\~n}oz, and K.~Pegion, 2022: {A decade of the North American Multimodel Ensemble (NMME): Research, application, and future directions}. \textit{Bull.\ Amer.\ Meteor.\ Soc.}, \textbf{103~(3)}, E973--E995, \doi{10.1175/BAMS-D-20-0327.1}.

\bibitem[{Bi et~al.(2023)Bi, Xie, Zhang, Chen, Gu,, and Tian}]{bi2023accurate}
Bi, K., L.~Xie, H.~Zhang, X.~Chen, X.~Gu, and Q.~Tian, 2023: Accurate medium-range global weather forecasting with 3d neural networks. \textit{Nature}, \textbf{619~(7970)}, 533--538, \doi{10.1038/s41586-023-06185-3}.

\bibitem[{Bjerknes(1969)}]{bjerknes1969atmospheric}
Bjerknes, J., 1969: {Atmospheric teleconnections from the equatorial Pacific}. \textit{Mon.\ Wea.\ Rev.}, \textbf{97~(3)}, 163--172, \doi{10.1175/1520-0493(1969)097<0163:ATFTEP>2.3.CO;2}.

\bibitem[{Blackmon(1976)}]{blackmon1976climatological}
Blackmon, M.~L., 1976: {A climatological spectral study of the 500 mb geopotential height of the Northern Hemisphere}. \textit{J.\ Atmos.\ Sci.}, \textbf{33~(8)}, 1607--1623, \doi{10.1175/1520-0469(1976)033<1607:ACSSOT>2.0.CO;2}.

\bibitem[{Büeler et~al.(2021)Büeler, Ferranti, Magnusson, Quinting,, and Grams}]{bueler2021}
Büeler, D., L.~Ferranti, L.~Magnusson, J.~F. Quinting, and C.~M. Grams, 2021: Year-round sub-seasonal forecast skill for atlantic–european weather regimes. \textit{Quarterly Journal of the Royal Meteorological Society}, \textbf{147~(741)}, 4283--4309, \doi{10.1002/qj.4178}, \urlprefix\url{https://rmets.onlinelibrary.wiley.com/doi/abs/10.1002/qj.4178}, \eprint{https://rmets.onlinelibrary.wiley.com/doi/pdf/10.1002/qj.4178}.

\bibitem[{Carton et~al.(2018)Carton, Chepurin,, and Chen}]{carton2018soda3}
Carton, J.~A., G.~A. Chepurin, and L.~Chen, 2018: {SODA3: A new ocean climate reanalysis}. \textit{J.\ Climate}, \textbf{31~(17)}, 6967--6983, \doi{10.1175/JCLI-D-18-0149.1}.

\bibitem[{Chelton et~al.(2011)Chelton, Schlax,, and Samelson}]{chelton2011global}
Chelton, D.~B., M.~G. Schlax, and R.~M. Samelson, 2011: Global observations of nonlinear mesoscale eddies. \textit{Progress in oceanography}, \textbf{91~(2)}, 167--216, \doi{10.1016/j.pocean.2011.01.002}.

\bibitem[{Chelton and Xie(2010)Chelton, and Xie}]{chelton2010coupled}
Chelton, D.~B., and S.-P. Xie, 2010: Coupled ocean-atmosphere interaction at oceanic mesoscales. \textit{Oceanography}, \textbf{23~(4)}, 52--69, \urlprefix\url{https://www.jstor.org/stable/24860862}.

\bibitem[{Chen et~al.(2024)}]{chen2024machine}
Chen, L., and Coauthors, 2024: A machine learning model that outperforms conventional global subseasonal forecast models. \textit{Nature Communications}, \textbf{15~(1)}, 6425, \doi{10.1038/s41467-024-50714-1}.

\bibitem[{Chen and Guestrin(2016)Chen, and Guestrin}]{chen2016xgboost}
Chen, T., and C.~Guestrin, 2016: {Xgboost: A scalable tree boosting system}. \textit{{Proceedings of the 22nd ACM SIGKDD International Conference on Knowledge Discovery and Data Mining}}, 785--794, \doi{10.1145/2939672.2939785}.

\bibitem[{Cheng and Wallace(1993)Cheng, and Wallace}]{cheng1993cluster}
Cheng, X., and J.~M. Wallace, 1993: {Cluster analysis of the Northern Hemisphere wintertime 500-hPa height field: Spatial patterns}. \textit{J.\ Atmos.\ Sci.}, \textbf{50~(16)}, 2674--2696, \doi{10.1175/1520-0469(1993)050<2674:CAOTNH>2.0.CO;2}.

\bibitem[{Conil et~al.(2009)Conil, Douville,, and Tyteca}]{conil2009contribution}
Conil, S., H.~Douville, and S.~Tyteca, 2009: Contribution of realistic soil moisture initial conditions to boreal summer climate predictability. \textit{Climate Dyn.}, \textbf{32}, 75--93, \doi{10.1007/s00382-008-0375-9}.

\bibitem[{Davini et~al.(2012)Davini, Cagnazzo, Neale,, and Tribbia}]{davini2012coupling}
Davini, P., C.~Cagnazzo, R.~Neale, and J.~Tribbia, 2012: {Coupling between Greenland blocking and the North Atlantic Oscillation pattern}. \textit{Geophys.\ Res.\ Lett.}, \textbf{39~(14)}, \doi{10.1029/2012GL052315}.

\bibitem[{DelSole et~al.(2017)DelSole, Trenary, Tippett,, and Pegion}]{PredictabilityofWeek34AverageTemperatureandPrecipitationovertheContiguousUnitedStates}
DelSole, T., L.~Trenary, M.~K. Tippett, and K.~Pegion, 2017: {Predictability of Week-3–4 Average Temperature and Precipitation over the Contiguous United States}. \textit{J.\ Climate}, \textbf{30~(10)}, 3499 -- 3512, \doi{10.1175/JCLI-D-16-0567.1}.

\bibitem[{Deser et~al.(2017)Deser, Simpson, McKinnon,, and Phillips}]{deser2017thenorthern}
Deser, C., I.~R. Simpson, K.~A. McKinnon, and A.~S. Phillips, 2017: The northern hemisphere extratropical atmospheric circulation response to enso: How well do we know it and how do we evaluate models accordingly? \textit{Journal of Climate}, \textbf{30~(13)}, 5059 -- 5082, \doi{10.1175/JCLI-D-16-0844.1}, \urlprefix\url{https://journals.ametsoc.org/view/journals/clim/30/13/jcli-d-16-0844.1.xml}.

\bibitem[{Deser et~al.(2007)Deser, Tomas,, and Peng}]{deser2007transient}
Deser, C., R.~A. Tomas, and S.~Peng, 2007: {The transient atmospheric circulation response to North Atlantic SST and sea ice anomalies}. \textit{J.\ Climate}, \textbf{20~(18)}, 4751--4767, \doi{10.1175/JCLI4278.1}.

\bibitem[{Domeisen et~al.(2020)}]{domeisen2020role}
Domeisen, D.~I., and Coauthors, 2020: {The role of the stratosphere in subseasonal to seasonal prediction: 2. Predictability arising from stratosphere-troposphere coupling}. \textit{Journal of Geophysical Research: Atmospheres}, \textbf{125~(2)}, e2019JD030\,923, \doi{10.1029/2019JD030923}.

\bibitem[{Dong et~al.(2023)Dong, Zeng, Wu, Huang, Gaiser,, and Srivastava}]{dong2023enhancing}
Dong, J., W.~Zeng, L.~Wu, J.~Huang, T.~Gaiser, and A.~K. Srivastava, 2023: {Enhancing short-term forecasting of daily precipitation using numerical weather prediction bias correcting with XGBoost in different regions of China}. \textit{Engineering Applications of Artificial Intelligence}, \textbf{117}, 105\,579, \doi{10.1016/j.engappai.2022.105579}.

\bibitem[{Fatima et~al.(2023)Fatima, Hussain, Amir, Ahmed,, and Aslam}]{fatima2023xgboost}
Fatima, S., A.~Hussain, S.~B. Amir, S.~H. Ahmed, and S.~M.~H. Aslam, 2023: {XGBoost and Random Forest Algorithms: An in Depth Analysis}. \textit{Pakistan Journal of Scientific Research}, \textbf{3~(1)}, 26--31, \doi{doi.org/10.57041/pjosr.v3i1.946}.

\bibitem[{Ferreira and Frankignoul(2005)Ferreira, and Frankignoul}]{ferreira2005transient}
Ferreira, D., and C.~Frankignoul, 2005: {The transient atmospheric response to midlatitude SST anomalies}. \textit{J.\ Climate}, \textbf{18~(7)}, 1049--1067, \doi{10.1175/JCLI-3313.1}.

\bibitem[{Frazier(2018)}]{frazier2018tutorialbayesianoptimization}
Frazier, P.~I., 2018: A tutorial on bayesian optimization. \urlprefix\url{https://arxiv.org/abs/1807.02811}, \doi{10.48550/arXiv.1807.02811}, \eprint{1807.02811}.

\bibitem[{Friedman(2001)}]{friedman2001greedy}
Friedman, J.~H., 2001: Greedy function approximation: A gradient boosting machine. \textit{{Annals of Statistics}}, 1189--1232, \urlprefix\url{https://www.jstor.org/stable/2699986}.

\bibitem[{Furtado et~al.(2012)Furtado, Di~Lorenzo, Anderson,, and Schneider}]{furtado2012linkages}
Furtado, J.~C., E.~Di~Lorenzo, B.~T. Anderson, and N.~Schneider, 2012: Linkages between the north pacific oscillation and central tropical pacific ssts at low frequencies. \textit{Climate Dynamics}, \textbf{39}, 2833--2846, \doi{10.1007/s00382-011-1245-4}.

\bibitem[{Gerber et~al.(2009)Gerber, Orbe,, and Polvani}]{gerber2009stratospheric}
Gerber, E., C.~Orbe, and L.~M. Polvani, 2009: Stratospheric influence on the tropospheric circulation revealed by idealized ensemble forecasts. \textit{Geophys.\ Res.\ Lett.}, \textbf{36~(24)}, \doi{10.1029/2009GL040913}.

\bibitem[{Gervais et~al.(2024)Gervais, Sun,, and Deser}]{gervais2024}
Gervais, M., L.~Sun, and C.~Deser, 2024: Impacts of projected arctic sea ice loss on daily weather patterns over north america. \textit{Journal of Climate}, \textbf{37~(3)}, 1065 -- 1085, \doi{10.1175/JCLI-D-23-0389.1}, \urlprefix\url{https://journals.ametsoc.org/view/journals/clim/37/3/JCLI-D-23-0389.1.xml}.

\bibitem[{Green and Furtado(2019)Green, and Furtado}]{green2019evaluating}
Green, M.~R., and J.~C. Furtado, 2019: Evaluating the joint influence of the madden-julian oscillation and the stratospheric polar vortex on weather patterns in the northern hemisphere. \textit{Journal of Geophysical Research: Atmospheres}, \textbf{124~(22)}, 11\,693--11\,709, \doi{10.1029/2019JD030771}, \urlprefix\url{https://agupubs.onlinelibrary.wiley.com/doi/abs/10.1029/2019JD030771}, \eprint{https://agupubs.onlinelibrary.wiley.com/doi/pdf/10.1029/2019JD030771}.

\bibitem[{Guo et~al.(2011)Guo, Dirmeyer,, and DelSole}]{guo2011land}
Guo, Z., P.~A. Dirmeyer, and T.~DelSole, 2011: Land surface impacts on subseasonal and seasonal predictability. \textit{Geophys.\ Res.\ Lett.}, \textbf{38~(24)}, \doi{10.1029/2011GL049945}.

\bibitem[{Guo et~al.(2012)Guo, Dirmeyer, DelSole,, and Koster}]{guo2012rebound}
Guo, Z., P.~A. Dirmeyer, T.~DelSole, and R.~D. Koster, 2012: Rebound in atmospheric predictability and the role of the land surface. \textit{J.\ Climate}, \textbf{25~(13)}, 4744--4749, \doi{10.1175/JCLI-D-11-00651.1}.

\bibitem[{Hartmann(2015)}]{hartmann2015pacific}
Hartmann, D.~L., 2015: Pacific sea surface temperature and the winter of 2014. \textit{Geophys.\ Res.\ Lett.}, \textbf{42~(6)}, 1894--1902, \doi{10.1002/2015GL063083}.

\bibitem[{Hasan et~al.(2020)Hasan, Zhen, Miah, Ahamed,, and Samie}]{hasan2020impact}
Hasan, S.~S., L.~Zhen, M.~G. Miah, T.~Ahamed, and A.~Samie, 2020: {Impact of land use change on ecosystem services: A review}. \textit{Environmental Development}, \textbf{34}, 100\,527, \doi{10.1016/j.envdev.2020.100527}.

\bibitem[{Hendon and Salby(1994)Hendon, and Salby}]{hendon1994life}
Hendon, H.~H., and M.~L. Salby, 1994: {The life cycle of the Madden--Julian oscillation}. \textit{J.\ Atmos.\ Sci.}, \textbf{51~(15)}, 2225--2237, \doi{10.1175/1520-0469(1994)051<2225:TLCOTM>2.0.CO;2}.

\bibitem[{Herman and Schumacher(2018)Herman, and Schumacher}]{herman2018money}
Herman, G.~R., and R.~S. Schumacher, 2018: {Money doesn't grow on trees, but forecasts do: Forecasting extreme precipitation with random forests}. \textit{Mon.\ Wea.\ Rev.}, \textbf{146~(5)}, 1571--1600, \doi{10.1175/MWR-D-17-0250.1}.

\bibitem[{Hersbach et~al.(2020)}]{hersbach2020era5}
Hersbach, H., and Coauthors, 2020: {The ERA5 global reanalysis}. \textit{Quart.\ J.\ Roy.\ Meteor.\ Soc.}, \textbf{146~(730)}, 1999--2049, \doi{10.1002/qj.3803}.

\bibitem[{Hochman et~al.(2021)Hochman, Messori, Quinting, Pinto,, and Grams}]{hochman2021}
Hochman, A., G.~Messori, J.~F. Quinting, J.~G. Pinto, and C.~M. Grams, 2021: Do atlantic-european weather regimes physically exist? \textit{Geophysical Research Letters}, \textbf{48~(20)}, e2021GL095\,574, \doi{10.1029/2021GL095574}, \urlprefix\url{https://agupubs.onlinelibrary.wiley.com/doi/abs/10.1029/2021GL095574}, e2021GL095574 2021GL095574, \eprint{https://agupubs.onlinelibrary.wiley.com/doi/pdf/10.1029/2021GL095574}.

\bibitem[{Horel and Wallace(1981)Horel, and Wallace}]{horel1981planetary}
Horel, J.~D., and J.~M. Wallace, 1981: {Planetary-scale atmospheric phenomena associated with the Southern Oscillation}. \textit{Mon.\ Wea.\ Rev.}, \textbf{109~(4)}, 813--829, \doi{10.1175/1520-0493(1981)109<0813:PSAPAW>2.0.CO;2}.

\bibitem[{Jennrich et~al.(2024)Jennrich, Straus, Chelliah,, and Baggett}]{jennrich2024}
Jennrich, G., D.~Straus, M.~Chelliah, and C.~Baggett, 2024: Ensemble predictability of week 3/4 precipitation and temperature over the united states via cluster analysis of the large-scale circulation. \textit{Weather and Forecasting}, \textbf{39~(11)}, 1531 -- 1544, \doi{10.1175/WAF-D-23-0065.1}, \urlprefix\url{https://journals.ametsoc.org/view/journals/wefo/39/11/WAF-D-23-0065.1.xml}.

\bibitem[{Jeong et~al.(2013)Jeong, Linderholm, Woo, Folland, Kim, Kim,, and Chen}]{jeong2013impacts}
Jeong, J.-H., H.~W. Linderholm, S.-H. Woo, C.~Folland, B.-M. Kim, S.-J. Kim, and D.~Chen, 2013: Impacts of snow initialization on subseasonal forecasts of surface air temperature for the cold season. \textit{J.\ Climate}, \textbf{26~(6)}, 1956--1972, \doi{10.1175/JCLI-D-12-00159.1}.

\bibitem[{Kidston et~al.(2015)Kidston, Scaife, Hardiman, Mitchell, Butchart, Baldwin,, and Gray}]{kidston2015stratospheric}
Kidston, J., A.~A. Scaife, S.~C. Hardiman, D.~M. Mitchell, N.~Butchart, M.~P. Baldwin, and L.~J. Gray, 2015: Stratospheric influence on tropospheric jet streams, storm tracks and surface weather. \textit{Nature Geoscience}, \textbf{8~(6)}, 433--440, \doi{10.1038/ngeo2424}.

\bibitem[{Kolstad et~al.(2022)Kolstad, Lee, Butler, Domeisen,, and Wulff}]{kolstad2022}
Kolstad, E.~W., S.~H. Lee, A.~H. Butler, D.~I.~V. Domeisen, and C.~O. Wulff, 2022: Diverse surface signatures of stratospheric polar vortex anomalies. \textit{Journal of Geophysical Research: Atmospheres}, \textbf{127~(20)}, e2022JD037\,422, \doi{10.1029/2022JD037422}, \urlprefix\url{https://agupubs.onlinelibrary.wiley.com/doi/abs/10.1029/2022JD037422}, e2022JD037422 2022JD037422, \eprint{https://agupubs.onlinelibrary.wiley.com/doi/pdf/10.1029/2022JD037422}.

\bibitem[{Korjus et~al.(2016)Korjus, Hebart,, and Vicente}]{korjus2016efficient}
Korjus, K., M.~N. Hebart, and R.~Vicente, 2016: An efficient data partitioning to improve classification performance while keeping parameters interpretable. \textit{PloS one}, \textbf{11~(8)}, e0161\,788, \doi{10.1371/journal.pone.0161788}.

\bibitem[{Koster and Walker(2015)Koster, and Walker}]{koster2015interactive}
Koster, R., and G.~Walker, 2015: Interactive vegetation phenology, soil moisture, and monthly temperature forecasts. \textit{J.\ Hydrometeor.}, \textbf{16~(4)}, 1456--1465, \doi{10.1175/JHM-D-14-0205.1}.

\bibitem[{Koster et~al.(2004)}]{koster2004regions}
Koster, R.~D., and Coauthors, 2004: Regions of strong coupling between soil moisture and precipitation. \textit{Science}, \textbf{305~(5687)}, 1138--1140, \doi{10.1126/science.1100217}.

\bibitem[{Koster et~al.(2010)}]{koster2010contribution}
Koster, R.~D., and Coauthors, 2010: {Contribution of land surface initialization to subseasonal forecast skill: First results from a multi-model experiment}. \textit{Geophys.\ Res.\ Lett.}, \textbf{37~(2)}, \doi{10.1029/2009GL041677}.

\bibitem[{Koster et~al.(2011)}]{TheSecondPhaseoftheGlobalLandAtmosphereCouplingExperimentSoilMoistureContributionstoSubseasonalForecastSkill}
Koster, R.~D., and Coauthors, 2011: {The Second Phase of the Global Land–Atmosphere Coupling Experiment: Soil Moisture Contributions to Subseasonal Forecast Skill}. \textit{J.\ Hydrometeor.}, \textbf{12~(5)}, 805 -- 822, \doi{https://doi.org/10.1175/2011JHM1365.1}.

\bibitem[{Krishnamurti(1961)}]{krishnamurti1961subtropical}
Krishnamurti, T.~N., 1961: The subtropical jet stream of winter. \textit{J.\ Atmos.\ Sci.}, \textbf{18~(2)}, 172--191, \doi{10.1175/1520-0469(1961)018<0172:TSJSOW>2.0.CO;2}.

\bibitem[{Kuang et~al.(2014)Kuang, Zhang, Huang,, and Huang}]{kuang2014spatial}
Kuang, X., Y.~Zhang, Y.~Huang, and D.~Huang, 2014: Spatial differences in seasonal variation of the upper-tropospheric jet stream in the northern hemisphere and its thermal dynamic mechanism. \textit{Theoretical and applied climatology}, \textbf{117}, 103--112, \doi{10.1007/s00704-013-0994-x}.

\bibitem[{Kurczyn et~al.(2012)Kurczyn, Beier, Lav{\'\i}n,, and Chaigneau}]{kurczyn2012mesoscale}
Kurczyn, J., E.~Beier, M.~F. Lav{\'\i}n, and A.~Chaigneau, 2012: {Mesoscale eddies in the northeastern Pacific tropical-subtropical transition zone: Statistical characterization from satellite altimetry}. \textit{Journal of Geophysical Research: Oceans}, \textbf{117~(C10)}, \doi{10.1029/2012JC007970}.

\bibitem[{Kushnir et~al.(2002)Kushnir, Robinson, Blad{\'e}, Hall, Peng,, and Sutton}]{kushnir2002atmospheric}
Kushnir, Y., W.~Robinson, I.~Blad{\'e}, N.~Hall, S.~Peng, and R.~Sutton, 2002: {Atmospheric GCM response to extratropical SST anomalies: Synthesis and evaluation}. \textit{J.\ Climate}, \textbf{15~(16)}, 2233--2256, \doi{10.1175/1520-0442(2002)015<2233:AGRTES>2.0.CO;2}.

\bibitem[{Kwon et~al.(2010)Kwon, Alexander, Bond, Frankignoul, Nakamura, Qiu,, and Thompson}]{kwon2010role}
Kwon, Y.-O., M.~A. Alexander, N.~A. Bond, C.~Frankignoul, H.~Nakamura, B.~Qiu, and L.~A. Thompson, 2010: {Role of the Gulf Stream and Kuroshio--Oyashio systems in large-scale atmosphere--ocean interaction: A review}. \textit{J.\ Climate}, \textbf{23~(12)}, 3249--3281, \doi{10.1175/2010JCLI3343.1}.

\bibitem[{Lang et~al.(2024)}]{lang2024aifsecmwfsdatadriven}
Lang, S., and Coauthors, 2024: Aifs -- ecmwf's data-driven forecasting system. \urlprefix\url{https://arxiv.org/abs/2406.01465}, \doi{10.48550/arXiv.2406.01465}, \eprint{2406.01465}.

\bibitem[{Lee et~al.(2019)Lee, Furtado,, and Charlton-Perez}]{lee2019wintertime}
Lee, S., J.~Furtado, and A.~Charlton-Perez, 2019: {Wintertime North American weather regimes and the Arctic stratospheric polar vortex}. \textit{Geophys.\ Res.\ Lett.}, \textbf{46~(24)}, 14\,892--14\,900, \doi{10.1029/2019GL085592}.

\bibitem[{Lee and Messori(2024)Lee, and Messori}]{lee2024}
Lee, S.~H., and G.~Messori, 2024: The dynamical footprint of year-round north american weather regimes. \textit{Geophysical Research Letters}, \textbf{51~(2)}, e2023GL107\,161, \doi{10.1029/2023GL107161}, \urlprefix\url{https://agupubs.onlinelibrary.wiley.com/doi/abs/10.1029/2023GL107161}, e2023GL107161 2023GL107161, \eprint{https://agupubs.onlinelibrary.wiley.com/doi/pdf/10.1029/2023GL107161}.

\bibitem[{Lee et~al.(2023)Lee, Tippett,, and Polvani}]{lee2023new}
Lee, S.~H., M.~K. Tippett, and L.~M. Polvani, 2023: {A new year-round weather regime classification for North America}. \textit{J.\ Climate}, \textbf{36~(20)}, 7091--7108, \doi{10.1175/JCLI-D-23-0214.1}.

\bibitem[{Lin et~al.(2019)Lin, Frederiksen, Straus,, and Stan}]{lin2019tropical}
Lin, H., J.~Frederiksen, D.~Straus, and C.~Stan, 2019: Tropical-extratropical interactions and teleconnections. \textit{Sub-Seasonal to Seasonal Prediction}, Elsevier, 143--164, \doi{10.1016/B978-0-12-811714-9.00007-3}.

\bibitem[{Lorenz(1963)}]{lorenz1963deterministic}
Lorenz, E.~N., 1963: Deterministic nonperiodic flow. \textit{J.\ Atmos.\ Sci.}, \textbf{20~(2)}, 130--141, \doi{10.1175/1520-0469(1963)020<0130:DNF>2.0.CO;2}.

\bibitem[{Lundberg and Lee(2017)Lundberg, and Lee}]{lundberg2017unifiedapproachinterpretingmodel}
Lundberg, S., and S.-I. Lee, 2017: A unified approach to interpreting model predictions. \urlprefix\url{https://arxiv.org/abs/1705.07874}, \doi{10.48550/arXiv.1705.07874}, \eprint{1705.07874}.

\bibitem[{L’Heureux and Thompson(2006)L’Heureux, and Thompson}]{lhereux2006observed}
L’Heureux, M.~L., and D.~W.~J. Thompson, 2006: Observed relationships between the el niño–southern oscillation and the extratropical zonal-mean circulation. \textit{Journal of Climate}, \textbf{19~(2)}, 276 -- 287, \doi{10.1175/JCLI3617.1}, \urlprefix\url{https://journals.ametsoc.org/view/journals/clim/19/2/jcli3617.1.xml}.

\bibitem[{Madden(1986)}]{madden1986seasonal}
Madden, R.~A., 1986: Seasonal variations of the 40-50 day oscillation in the tropics. \textit{Journal of Atmospheric Sciences}, \textbf{43~(24)}, 3138--3158.

\bibitem[{Mariotti et~al.(2020)}]{mariotti2020windows}
Mariotti, A., and Coauthors, 2020: Windows of opportunity for skillful forecasts subseasonal to seasonal and beyond. \textit{Bull.\ Amer.\ Meteor.\ Soc.}, \textbf{101~(5)}, E608--E625, \doi{10.1175/BAMS-D-18-0326.1}.

\bibitem[{Mayer and Barnes(2021)Mayer, and Barnes}]{mayer2021subseasonal}
Mayer, K.~J., and E.~A. Barnes, 2021: Subseasonal forecasts of opportunity identified by an explainable neural network. \textit{Geophys.\ Res.\ Lett.}, \textbf{48~(10)}, e2020GL092\,092, \doi{10.1029/2020GL092092}.

\bibitem[{Mayer et~al.(2024)Mayer, Chapman,, and Manriquez}]{mayer2024}
Mayer, K.~J., W.~E. Chapman, and W.~A. Manriquez, 2024: Exploring the relative importance of the mjo and enso to north pacific subseasonal predictability. \textit{Geophysical Research Letters}, \textbf{51~(10)}, e2024GL108\,479, \doi{10.1029/2024GL108479}, \urlprefix\url{https://agupubs.onlinelibrary.wiley.com/doi/abs/10.1029/2024GL108479}, e2024GL108479 2024GL108479, \eprint{https://agupubs.onlinelibrary.wiley.com/doi/pdf/10.1029/2024GL108479}.

\bibitem[{Meehl et~al.(2021)}]{meehl2021initialized}
Meehl, G.~A., and Coauthors, 2021: {Initialized Earth system prediction from subseasonal to decadal timescales}. \textit{Nature Reviews Earth \& Environment}, \textbf{2~(5)}, 340--357, \doi{10.1038/s43017-021-00155-x}.

\bibitem[{Merryfield et~al.(2020)}]{merryfield2020current}
Merryfield, W.~J., and Coauthors, 2020: Current and emerging developments in subseasonal to decadal prediction. \textit{Bull.\ Amer.\ Meteor.\ Soc.}, \textbf{101~(6)}, E869--E896, \doi{10.1175/BAMS-D-19-0037.1}.

\bibitem[{Michelangeli et~al.(1995)Michelangeli, Vautard,, and Legras}]{michelangeli1995weather}
Michelangeli, P.-A., R.~Vautard, and B.~Legras, 1995: Weather regimes: Recurrence and quasi stationarity. \textit{J.\ Atmos.\ Sci.}, \textbf{52~(8)}, 1237--1256, \doi{10.1175/1520-0469(1995)052<1237:WRRAQS>2.0.CO;2}.

\bibitem[{Molina et~al.(2023{\natexlab{a}})Molina, Richter, Glanville, Dagon, Berner, Hu,, and Meehl}]{molina2023subseasonal}
Molina, M.~J., J.~H. Richter, A.~A. Glanville, K.~Dagon, J.~Berner, A.~Hu, and G.~A. Meehl, 2023{\natexlab{a}}: {Subseasonal representation and predictability of North American weather regimes using cluster analysis}. \textit{Artificial Intelligence for the Earth Systems}, 1--54, \doi{10.1175/AIES-D-22-0051.1}.

\bibitem[{Molina et~al.(2023{\natexlab{b}})}]{molina2023review}
Molina, M.~J., and Coauthors, 2023{\natexlab{b}}: A review of recent and emerging machine learning applications for climate variability and weather phenomena. \textit{Artificial Intelligence for the Earth Systems}, 1--46, \doi{10.1175/AIES-D-22-0086.1}.

\bibitem[{Mostajabi et~al.(2019)Mostajabi, Finney, Rubinstein,, and Rachidi}]{mostajabi2019nowcasting}
Mostajabi, A., D.~L. Finney, M.~Rubinstein, and F.~Rachidi, 2019: Nowcasting lightning occurrence from commonly available meteorological parameters using machine learning techniques. \textit{Npj Climate and Atmospheric Science}, \textbf{2~(1)}, 41, \doi{10.1038/s41612-019-0098-0}.

\bibitem[{Mu{\~n}oz-Sabater et~al.(2021)}]{munoz2021era5}
Mu{\~n}oz-Sabater, J., and Coauthors, 2021: {ERA5-Land: A state-of-the-art global reanalysis dataset for land applications}. \textit{Earth System Science Data}, \textbf{13~(9)}, 4349--4383, \doi{10.5194/essd-13-4349-2021}.

\bibitem[{National Academies~of Sciences et~al.(2016)National Academies~of Sciences, Medicine, Climate,, and Studies}]{board2016next}
National Academies~of Sciences, E., B.~o. A.~S. Medicine, D.~o.~E. Climate, Ocean Studies~Board, and L.~Studies, 2016: \textit{{Next generation Earth system prediction: Strategies for subseasonal to seasonal forecasts}}. National Academies Press, Washington, D.C., \doi{10.17226/21873}.

\bibitem[{Orsolini et~al.(2013)Orsolini, Senan, Balsamo, Doblas-Reyes, Vitart, Weisheimer, Carrasco,, and Benestad}]{orsolini2013impact}
Orsolini, Y., R.~Senan, G.~Balsamo, F.~Doblas-Reyes, F.~Vitart, A.~Weisheimer, A.~Carrasco, and R.~Benestad, 2013: Impact of snow initialization on sub-seasonal forecasts. \textit{Climate Dyn.}, \textbf{41}, 1969--1982, \doi{10.1007/s00382-013-1782-0}.

\bibitem[{Palm{\'e}n(1948)}]{palmen1948distribution}
Palm{\'e}n, E., 1948: On the distribution of temperature and wind in the upper westerlies. \textit{J.\ Atmos.\ Sci.}, \textbf{5~(1)}, 20--27, \doi{10.1175/1520-0469(1948)005<0020:OTDOTA>2.0.CO;2}.

\bibitem[{Parker et~al.(2018)Parker, Woollings,, and Weisheimer}]{parker2018ensemble}
Parker, T., T.~Woollings, and A.~Weisheimer, 2018: Ensemble sensitivity analysis of greenland blocking in medium-range forecasts. \textit{Quarterly Journal of the Royal Meteorological Society}, \textbf{144~(716)}, 2358--2379, \doi{10.1002/qj.3391}, \urlprefix\url{https://rmets.onlinelibrary.wiley.com/doi/abs/10.1002/qj.3391}, \eprint{https://rmets.onlinelibrary.wiley.com/doi/pdf/10.1002/qj.3391}.

\bibitem[{Pegion et~al.(2019)}]{pegion2019subseasonal}
Pegion, K., and Coauthors, 2019: {The Subseasonal Experiment (SubX): A multimodel subseasonal prediction experiment}. \textit{Bull.\ Amer.\ Meteor.\ Soc.}, \textbf{100~(10)}, 2043--2060, \doi{10.1175/BAMS-D-18-0270.1}.

\bibitem[{Peings et~al.(2011)Peings, Douville, Alkama,, and Decharme}]{peings2011snow}
Peings, Y., H.~Douville, R.~Alkama, and B.~Decharme, 2011: Snow contribution to springtime atmospheric predictability over the second half of the twentieth century. \textit{Climate Dyn.}, \textbf{37}, 985--1004, \doi{10.1007/s00382-010-0884-1}.

\bibitem[{Petrie et~al.(2015)Petrie, Shaffrey,, and Sutton}]{petrieetal2015}
Petrie, R.~E., L.~C. Shaffrey, and R.~T. Sutton, 2015: Atmospheric response in summer linked to recent arctic sea ice loss. \textit{Quarterly Journal of the Royal Meteorological Society}, \textbf{141~(691)}, 2070--2076, \doi{10.1002/qj.2502}, \urlprefix\url{https://rmets.onlinelibrary.wiley.com/doi/abs/10.1002/qj.2502}, \eprint{https://rmets.onlinelibrary.wiley.com/doi/pdf/10.1002/qj.2502}.

\bibitem[{Philander(1999)}]{philander1999review}
Philander, S.~G., 1999: A review of tropical ocean--atmosphere interactions. \textit{Tellus B}, \textbf{51~(1)}, 71--90.

\bibitem[{Philander(1983)}]{philander1983nino}
Philander, S. G.~H., 1983: {El Niño--Southern Oscillation phenomena}. \textit{Nature}, \textbf{302~(5906)}, 295--301, \doi{10.1038/302295a0}.

\bibitem[{Qian et~al.(2020)Qian, Jia,, and Lin}]{qian2020machine}
Qian, Q.~F., X.~J. Jia, and H.~Lin, 2020: {Machine learning models for the seasonal forecast of winter surface air temperature in North America}. \textit{Earth and Space Science}, \textbf{7~(8)}, e2020EA001\,140, \doi{10.1029/2020EA001140}.

\bibitem[{Reinhold and Pierrehumbert(1982)Reinhold, and Pierrehumbert}]{reinhold1982dynamics}
Reinhold, B.~B., and R.~T. Pierrehumbert, 1982: {Dynamics of weather regimes: Quasi-stationary waves and blocking}. \textit{Mon.\ Wea.\ Rev.}, \textbf{110~(9)}, 1105--1145, \doi{10.1175/1520-0493(1982)110<1105:DOWRQS>2.0.CO;2}.

\bibitem[{Richter et~al.(2024)}]{richter2024quantifying}
Richter, J.~H., and Coauthors, 2024: {Quantifying sources of subseasonal prediction skill in CESM2}. \textit{npj Climate and Atmospheric Science}, \textbf{7~(1)}, 59, \doi{10.1038/s41612-024-00595-4}.

\bibitem[{Robertson and Vitart(2018)Robertson, and Vitart}]{robertson2018sub}
Robertson, A., and F.~Vitart, 2018: \textit{Sub-seasonal to seasonal prediction: the gap between weather and climate forecasting}. Elsevier.

\bibitem[{Robertson et~al.(2020)Robertson, Vigaud, Yuan,, and Tippett}]{robertson2020toward}
Robertson, A.~W., N.~Vigaud, J.~Yuan, and M.~K. Tippett, 2020: {Toward identifying subseasonal forecasts of opportunity using North American weather regimes}. \textit{Mon.\ Wea.\ Rev.}, \textbf{148~(5)}, 1861--1875, \doi{10.1175/MWR-D-19-0285.1}.

\bibitem[{Roundy and Wood(2015)Roundy, and Wood}]{roundy2015attribution}
Roundy, J.~K., and E.~F. Wood, 2015: The attribution of land--atmosphere interactions on the seasonal predictability of drought. \textit{J.\ Hydrometeor.}, \textbf{16~(2)}, 793--810, \doi{10.1175/JHM-D-14-0121.1}.

\bibitem[{Saji et~al.(1999)Saji, Goswami, Vinayachandran,, and Yamagata}]{saji1999dipole}
Saji, N., B.~N. Goswami, P.~Vinayachandran, and T.~Yamagata, 1999: {A dipole mode in the tropical Indian Ocean}. \textit{Nature}, \textbf{401~(6751)}, 360--363, \doi{10.1038/43854}.

\bibitem[{Scherrer et~al.(2006)Scherrer, Croci-Maspoli, Schwierz,, and Appenzeller}]{scherrer2006two}
Scherrer, S.~C., M.~Croci-Maspoli, C.~Schwierz, and C.~Appenzeller, 2006: {Two-dimensional indices of atmospheric blocking and their statistical relationship with winter climate patterns in the Euro-Atlantic region}. \textit{International Journal of Climatology: A Journal of the Royal Meteorological Society}, \textbf{26~(2)}, 233--249, \doi{10.1002/joc.1250}.

\bibitem[{Schmitt(2022)}]{schmitt2022deep}
Schmitt, M., 2022: Deep learning vs. gradient boosting: Benchmarking state-of-the-art machine learning algorithms for credit scoring. \textit{arXiv preprint arXiv:2205.10535}, \doi{10.48550/arXiv.2205.10535}.

\bibitem[{Screen et~al.(2011)Screen, Simmonds,, and Keay}]{screen2011dramatic}
Screen, J.~A., I.~Simmonds, and K.~Keay, 2011: {Dramatic interannual changes of perennial Arctic sea ice linked to abnormal summer storm activity}. \textit{Journal of Geophysical Research: Atmospheres}, \textbf{116~(D15)}, \doi{10.1029/2011JD015847}.

\bibitem[{Sengupta et~al.(2022)Sengupta, Singh, DeFlorio, Raymond, Robertson, Zeng, Waliser,, and Jones}]{sengupta2022advances}
Sengupta, A., B.~Singh, M.~J. DeFlorio, C.~Raymond, A.~W. Robertson, X.~Zeng, D.~E. Waliser, and J.~Jones, 2022: {Advances in subseasonal to seasonal prediction relevant to water management in the western United States}. \textit{Bull.\ Amer.\ Meteor.\ Soc.}, \textbf{103~(10)}, E2168--E2175, \doi{10.1175/BAMS-D-22-0146.1}.

\bibitem[{Snoek et~al.(2012)Snoek, Larochelle,, and Adams}]{snoek2012practical}
Snoek, J., H.~Larochelle, and R.~P. Adams, 2012: {Practical Bayesian optimization of machine learning algorithms}. \textit{Advances in neural information processing systems}, \textbf{25}.

\bibitem[{Stan et~al.(2017)Stan, Straus, Frederiksen, Lin, Maloney,, and Schumacher}]{stan2017review}
Stan, C., D.~M. Straus, J.~S. Frederiksen, H.~Lin, E.~D. Maloney, and C.~Schumacher, 2017: Review of tropical-extratropical teleconnections on intraseasonal time scales. \textit{Rev. Geophys.}, \textbf{55~(4)}, 902--937, \doi{10.1002/2016RG000538}.

\bibitem[{Straus et~al.(2007)Straus, Corti,, and Molteni}]{straus2007}
Straus, D.~M., S.~Corti, and F.~Molteni, 2007: Circulation regimes: Chaotic variability versus sst-forced predictability. \textit{Journal of Climate}, \textbf{20~(10)}, 2251 -- 2272, \doi{10.1175/JCLI4070.1}, \urlprefix\url{https://journals.ametsoc.org/view/journals/clim/20/10/jcli4070.1.xml}.

\bibitem[{Strong and Davis(2007)Strong, and Davis}]{strong2007winter}
Strong, C., and R.~E. Davis, 2007: {Winter jet stream trends over the Northern Hemisphere}. \textit{Quart.\ J.\ Roy.\ Meteor.\ Soc.}, \textbf{133~(629)}, 2109--2115, \doi{10.1002/qj.171}.

\bibitem[{Teng et~al.(2019)Teng, Branstator, Tawfik,, and Callaghan}]{CircumglobalResponsetoPrescribedSoilMoistureoverNorthAmerica}
Teng, H., G.~Branstator, A.~B. Tawfik, and P.~Callaghan, 2019: {Circumglobal response to prescribed soil moisture over North America}. \textit{J.\ Climate}, \textbf{32~(14)}, 4525 -- 4545, \doi{10.1175/JCLI-D-18-0823.1}.

\bibitem[{Thomas et~al.(2016)Thomas, Berg,, and Merryfield}]{thomas2016influence}
Thomas, J.~A., A.~A. Berg, and W.~J. Merryfield, 2016: Influence of snow and soil moisture initialization on sub-seasonal predictability and forecast skill in boreal spring. \textit{Climate Dyn.}, \textbf{47}, 49--65, \doi{10.1007/s00382-015-2821-9}.

\bibitem[{Thompson and Wallace(2000)Thompson, and Wallace}]{thompson2000annular}
Thompson, D.~W., and J.~M. Wallace, 2000: {Annular modes in the extratropical circulation. Part I: Month-to-month variability}. \textit{J.\ Climate}, \textbf{13~(5)}, 1000--1016, \doi{10.1175/1520-0442(2000)013<1000:AMITEC>2.0.CO;2}.

\bibitem[{Tibaldi and Molteni(1990)Tibaldi, and Molteni}]{tibaldi1990operational}
Tibaldi, S., and F.~Molteni, 1990: On the operational predictability of blocking. \textit{Tellus A}, \textbf{42~(3)}, 343--365, \doi{10.1034/j.1600-0870.1990.t01-2-00003.x}.

\bibitem[{Trenberth et~al.(1998)Trenberth, Branstator, Karoly, Kumar, Lau,, and Ropelewski}]{trenberth1998progress}
Trenberth, K.~E., G.~W. Branstator, D.~Karoly, A.~Kumar, N.-C. Lau, and C.~Ropelewski, 1998: {Progress during TOGA in understanding and modeling global teleconnections associated with tropical sea surface temperatures}. \textit{Journal of Geophysical Research: Oceans}, \textbf{103~(C7)}, 14\,291--14\,324, \doi{10.1029/97JC01444}.

\bibitem[{Tseng et~al.(2018)Tseng, Barnes,, and Maloney}]{tseng2018prediction}
Tseng, K.-C., E.~Barnes, and E.~Maloney, 2018: {Prediction of the midlatitude response to strong Madden-Julian Oscillation events on S2S time scales}. \textit{Geophys.\ Res.\ Lett.}, \textbf{45~(1)}, 463--470, \doi{10.1002/2017GL075734}.

\bibitem[{Uchoa et~al.(2023)Uchoa, Simoes-Sousa,, and da~Silveira}]{uchoa2023brazil}
Uchoa, I., I.~T. Simoes-Sousa, and I.~C. da~Silveira, 2023: {The Brazil Current mesoscale eddies: Altimetry-based characterization and tracking}. \textit{Deep Sea Research Part I: Oceanographic Research Papers}, \textbf{192}, 103\,947, \doi{10.1016/j.dsr.2022.103947}.

\bibitem[{Vigaud et~al.(2018)Vigaud, Robertson,, and Tippett}]{vigaud2018predictability}
Vigaud, N., A.~W. Robertson, and M.~K. Tippett, 2018: {Predictability of recurrent weather regimes over North America during winter from submonthly reforecasts}. \textit{Mon.\ Wea.\ Rev.}, \textbf{146~(8)}, 2559--2577, \doi{10.1175/MWR-D-18-0058.1}.

\bibitem[{Vitart et~al.(2012)Vitart, Robertson,, and Anderson}]{vitart2012subseasonal}
Vitart, F., A.~W. Robertson, and D.~L. Anderson, 2012: Subseasonal to seasonal prediction project: Bridging the gap between weather and climate. \textit{Bulletin of the World Meteorological Organization}, \textbf{61~(2)}, 23.

\bibitem[{Vitart et~al.(2022)}]{vitart2022outcomes}
Vitart, F., and Coauthors, 2022: {Outcomes of the WMO Prize Challenge to improve subseasonal to seasonal predictions using artificial intelligence}. \textit{Bull.\ Amer.\ Meteor.\ Soc.}, \textbf{103~(12)}, E2878--E2886, \doi{10.1175/BAMS-D-22-0046.1}.

\bibitem[{Wallace and Gutzler(1981)Wallace, and Gutzler}]{wallace1981teleconnections}
Wallace, J.~M., and D.~S. Gutzler, 1981: {Teleconnections in the geopotential height field during the Northern Hemisphere winter}. \textit{Mon.\ Wea.\ Rev.}, \textbf{109~(4)}, 784--812, \doi{10.1175/1520-0493(1981)109<0784:TITGHF>2.0.CO;2}.

\bibitem[{Wang et~al.(2018)Wang, Tian, Xie, Zhang,, and Han}]{wang2018mjo}
Wang, F., W.~Tian, F.~Xie, J.~Zhang, and Y.~Han, 2018: Effect of madden–julian oscillation occurrence frequency on the interannual variability of northern hemisphere stratospheric wave activity in winter. \textit{Journal of Climate}, \textbf{31~(13)}, 5031 -- 5049, \doi{10.1175/JCLI-D-17-0476.1}, \urlprefix\url{https://journals.ametsoc.org/view/journals/clim/31/13/jcli-d-17-0476.1.xml}.

\bibitem[{Weyn et~al.(2019)Weyn, Durran,, and Caruana}]{weyn2021can}
Weyn, J.~A., D.~R. Durran, and R.~Caruana, 2019: Can machines learn to predict weather? using deep learning to predict gridded 500-hpa geopotential height from historical weather data. \textit{Journal of Advances in Modeling Earth Systems}, \textbf{11~(8)}, 2680--2693, \doi{https://doi.org/10.1029/2019MS001705}, \urlprefix\url{https://agupubs.onlinelibrary.wiley.com/doi/abs/10.1029/2019MS001705}, \eprint{https://agupubs.onlinelibrary.wiley.com/doi/pdf/10.1029/2019MS001705}.

\bibitem[{Weyn et~al.(2021)Weyn, Durran, Caruana,, and Cresswell-Clay}]{weyn2021sub}
Weyn, J.~A., D.~R. Durran, R.~Caruana, and N.~Cresswell-Clay, 2021: Sub-seasonal forecasting with a large ensemble of deep-learning weather prediction models. \textit{Journal of Advances in Modeling Earth Systems}, \textbf{13~(7)}, e2021MS002\,502, \doi{10.1029/2021MS002502}.

\bibitem[{Wheeler and Kiladis(1999)Wheeler, and Kiladis}]{wheeler1999convectively}
Wheeler, M., and G.~N. Kiladis, 1999: {Convectively coupled equatorial waves: Analysis of clouds and temperature in the wavenumber--frequency domain}. \textit{J.\ Atmos.\ Sci.}, \textbf{56~(3)}, 374--399, \doi{10.1175/1520-0469(1999)056<0374:CCEWAO>2.0.CO;2}.

\bibitem[{Wheeler and Hendon(2004)Wheeler, and Hendon}]{wheeler2004all}
Wheeler, M.~C., and H.~H. Hendon, 2004: {An all-season real-time multivariate MJO index: Development of an index for monitoring and prediction}. \textit{Mon.\ Wea.\ Rev.}, \textbf{132~(8)}, 1917--1932, \doi{10.1175/1520-0493(2004)132<1917:AARMMI>2.0.CO;2}.

\bibitem[{White et~al.(2017)}]{white2017potential}
White, C.~J., and Coauthors, 2017: {Potential applications of subseasonal-to-seasonal (S2S) predictions}. \textit{Meteorological Applications}, \textbf{24~(3)}, 315--325, \doi{10.1002/met.1654}.

\bibitem[{Yulaeva and Wallace(1994)Yulaeva, and Wallace}]{yulaeva1994signature}
Yulaeva, E., and J.~M. Wallace, 1994: {The signature of ENSO in global temperature and precipitation fields derived from the microwave sounding unit}. \textit{J.\ Climate}, \textbf{7~(11)}, 1719--1736, \doi{10.1175/1520-0442(1994)007<1719:TSOEIG>2.0.CO;2}.

\bibitem[{Zhou(2019)}]{zhou2019atmospheric}
Zhou, G., 2019: {Atmospheric response to sea surface temperature anomalies in the mid-latitude oceans: A brief review}. \textit{Atmosphere-Ocean}, \textbf{57~(5)}, 319--328, \doi{10.1080/07055900.2019.1702499}.

\end{thebibliography}

\end{document}